\documentclass[aip,amsmath,amssymb,reprint]{revtex4-2}
\usepackage{amssymb}
\usepackage{amsmath}
\usepackage{color}
\usepackage[utf8]{inputenc}
\usepackage{graphicx}
\usepackage{epstopdf}

\usepackage{hyperref}

\newcommand{\be}{\begin{equation}}
\newcommand{\ee}{\end{equation}}
\newcommand{\bea}{\begin{eqnarray}}
\newcommand{\eea}{\end{eqnarray}}

\newcommand{\beginsupplement}{
\setcounter{table}{0}
\renewcommand{\thetable}{S\arabic{table}}
\setcounter{figure}{0}
\renewcommand{\thefigure}{S\arabic{figure}}
\setcounter{equation}{0}
\renewcommand{\theequation}{S\arabic{equation}}
}

\begin{document}

\title{Mapping the influence of impurity interaction energy on nucleation in a lattice-gas model of solute precipitation}
 
\author{Dipanjan Mandal}
\email{dipanjan.mandal@warwick.ac.uk}
\affiliation{Department of Physics, University of Warwick,Coventry,CV4 7AL,United Kingdom}

\author{David Quigley}
\email{d.quigley@warwick.ac.uk}
\affiliation{Department of Physics, University of Warwick,Coventry,CV4 7AL,United Kingdom}

\date{\today}

\begin{abstract}
We study nucleation in the two dimensional Ising lattice-gas model of solute precipitation in the presence of randomly placed static and dynamic impurities. Impurity-solute and impurity-solvent interaction energies are varied whilst keeping other interaction energies fixed. In the case of static impurities, we observe a monotonic decrease in the nucleation rate when the difference between impurity-solute and impurity-solvent interaction energies is increased. The nucleation rate saturates to a minimum value with increasing interaction energy difference when the impurity density is low. However the nucleation rate does not saturate for high impurity densities. Similar behaviour is observed with dynamic impurities both at low and high densities. We explore a broad range of both symmetric and anti-symmetric interactions with impurities and map the regime for which the impurities act as a surfactant, decreasing the surface energy of the nucleating phase. We also characterise different nucleation regimes observed at different values of interaction energy. These include additional regimes where impurities play the role of inert-spectators, bulk-stabilizers or cluster together to create heterogeneous nucleation sites for solute clusters to form.
\end{abstract}

\maketitle

\section{Introduction}
\label{intro}
Nucleation is the mechanism by which a stable phase emerges from a metastable parent phase. It is the first step in the synthesis of many materials and frequently observed in nature. Classical nucleation theory~\cite{1935-becker-aphys,zeldovich,Kashchiev2000} (CNT) is a well-accepted theory which can quantitatively explain this mechanism provided certain assumptions hold. A simple system of particles with short range attractive or repulsive interactions, like the 2D Ising lattice gas, exhibits nucleation behaviour accurately predicted from CNT via the Becker-Doring-Zeldovich expression~\cite{cai2010pre}. This model has been used to test assumptions and predictions of CNT in several previous studies~\cite{mc-prl-1982,cnt-jcp-1999,mc-pre-2013-binder}. 

In the context of solute precipitation, occupied lattice sites in this model are considered as solute particles, with empty lattice sites representing solvent. The nucleating phase transition is hence from a supersaturated solution to a precipitated 
solid phase. Interpreted in this way,  lattice models have been used to capture two-step nucleation mechanisms~\cite{duff_jcp_2009} and more complex nucleation pathways~\cite{lifanov_jcp_2016}.

Impurities or additives are often used to control the nucleation process~\cite{TEYCHENE20201}, either accelerating~\cite{2018_surfactant} or decelerating~\cite{additive_caco3_2009,Xu2024_additive} the nucleation rate. However, the mechanism by which additives influence the cluster growth varies. For example, impurities may capture the solute ions~\cite{additive_caco3_2009} or inhibit or enhance crystal growth via attachment to the surface~\cite{2019_additive_surface,2018_surfactant,Xu2024_additive}. A common real-world example where this surfactant property is used as a cleansing mechanism is the formation of micelles in aqueous solutions of soap. In recent studies, the role of structure and size of impurity particles in determining the surface properties as well as the nucleation rate is explored using Molecular Dynamics simulations~\cite{anwar_2009,anwar_2011}. It has been shown that particle size asymmetry inhibit the nucleation in spite of strong affinity between solute and impurity. In the case of polymeric impurities, the interfacial tension is decreased as the impurities bind at the interface~\cite{bertolazzo2018}. The chain length of the polymer also plays significant role in nucleation. The nucleation free energy barrier is decreased with increasing the chain length of the polymer~\cite{poon2017}. Presence of ionic impurities in aqueous nano-droplet lead to demixing of low density amorphous ice and impurity rich aqueous glass when cooled~\cite{hudait2014}.

Although, CNT was constructed for homogeneous nucleation in pure systems, we recently demonstrated that nucleation rates for a system containing a low concentration of homogeneously distributed impurities can be predicted by simple modification of the interfacial energy~\cite{mandal2021sm} term. In Ref.~\cite{mandal2021sm} a special case, where both solute and solvent particles exhibit energy neutral interactions with impurities, was considered. In this scenario the impurities act as a surfactant, lowering the interfacial free energy between nuclei and the parent phase. 

Studies of simple lattice models are useful to gain insight into complex nucleation mechanisms in many scenarios.  Previous studies have examined the effects of surface pore-width~\cite{page2006prl}, pore-geometry~\cite{whitelam2012sm} and surface roughness~\cite{lutsko2021}
on nucleation rate in the Ising model. Intelligent choice of surface geometry certainly helps to optimize the synthesis of new materials from the solution of its constituent particles. Studying competing nucleation between stable vs. metastable precipitated phases in a dimer lattice gas~\cite{2022_mandal} provides insight into multi-component nucleation mechanism. Furthermore, examining the role of defects on magnetic droplet nucleation~\cite{ettori_2023}, studying nucleation in the random field Ising model~\cite{yao_heterogeneous} and Potts lattice-gas model~\cite{peters_additive_2015,poon2017} are other examples where simple lattice models are have captured emergent nucleation behaviour. We note that in the Potts lattice gas model~\cite{peters_additive_2015}, a surfactant-like additive was used to enhance the nucleation rate by decreasing the surface tension, similar to our previous work~~\cite{mandal2021sm}. 

In this paper we explore a wide range of alternative scenarios, in which solute and solvent particles exhibit equal (symmetric) or opposite (anti-symmetric) interactions with impurities, over a wide range of interaction energies. We demonstrate that the surfactant-like behaviour reported previously is just one of several mechanisms by which impurities modify nucleation that can be captured within these minimal models.

The remainder of the paper is organised as follows. In Section~\ref{sec:model}, we describe the model and the algorithm used for the simulations. The impact of impurity interactions on free energy barriers to nucleation is shown in Section~\ref{sec:static} and Section~\ref{sec:dynamic} for static and dynamic impurities respectively. We characterize the different nucleation regimes that emerge from interaction with mobile impurities in Section~\ref{sec:dynamic}. Section~\ref{sec:rate} contains results on nucleation rate obtained directly from forward flux sampling, which are compared to predictions made by CNT. A particular regime which exhibits clustering of impurities and enabling cross-nucleation of the stable phase is described in Section~\ref{sec:imp_clustering}. Finally, we conclude in Section~\ref{sec:conclusion}.

\section{Model \& Algorithm}
\label{sec:model}
We consider a two dimensional Ising gas in the presence of randomly positioned impurities on an $L\times L$ square lattice  where $L=100$. Each site has a variable $s_i$ denoting the occupancy of site $i$ which can be either solute, solvent or impurity particles which symbolised by $s_i = u$, $v$ or $i$ respectively. 
We work in the semi-grand ensemble, expressing the corresponding potential as
\be
\label{eq:energy}
\Phi=\sum_{\langle i,j\rangle}\epsilon_{s_i, s_j}-\frac{\Delta \mu}{2}\sum_i(\delta_{s_i,u}-\delta_{s_i,v})
\ee
where $\epsilon_{s_i, s_j}$ is the interaction energy between the species $s_i$ and $s_j$ when $i$ and $j$ are nearest neighbours, and $\Delta \mu = \mu_u - \mu_v$ is the chemical potential difference between a solute reservoir and solvent reservoir with which the system can exchange particles. In matrix form the interaction energies can be written as:
\be
\epsilon_{s_i, s_j}=
\begin{bmatrix}
    -J       & J & \epsilon_+ \\
    J       & -J & \epsilon_-  \\
    \epsilon_+       & \epsilon_- & 0 
\end{bmatrix},
\ee
where the indices are $s_i, s_j=u, v, i$, starting from top left corner of the matrix. The upper-left $2 \times 2$ sub-matrix is the familiar Ising lattice gas with nearest neighbour interaction energy $J$. The remaining terms represent coupling between solute/solvent and impurity particles.
We simulate below the critical temperature, such that any positive value of $\Delta \mu$ correspond to conditions in which the solution is supersaturated and metastable with respect to the formation of a solute rich phase.

The interaction energies of solute and solvent with impurity are denoted by $\epsilon_{u,i}=\epsilon_+$ and $\epsilon_{v,i}=\epsilon_-$ respectively throughout the paper. We study the nucleation behaviour for different values of these interaction energies at fixed impurity density. These can be categorised into three groups which are symmetric ($\epsilon_+=\epsilon_-$), anti-symmetric ($\epsilon_+=-\epsilon_-$) and asymmetric ($|\epsilon_+|\neq |\epsilon_-|$) interaction energy. 
We set the strength of the interaction energy coupling $J=1$, both within and between solute and solvent, and the impurity-impurity interaction strength is set to $0$ throughout the models studied in the paper. 

We simulate this model using Metropolis Monte Carlo dynamics. The usual transmutation moves of solute into solvent and vice-versa represent removal of one species to its corresponding reservoir and replacement from the other reservoir. 

In addition we introduce moves which allow the impurities to migrate. We define a mobility parameter $\alpha$ of the impurities such that $\alpha=0$ corresponds to static impurities and $\alpha=1$ corresponds to the fastest moving impurities with no transmutation dynamics. At each Monte Carlo move we randomly generate a random number $\xi$ uniformly distributed between 0 and 1. If $\xi<\alpha$, we attempt a "non-local swap" move in which the occupancy of a randomly selected impurity site is swapped with a second randomly selected site a distance $d$ away. Specifically we generate a displacement vector $(\Delta x, \Delta y)$ with circular symmetry by setting $\Delta x$ to a random integer between $-d$ to $d$ then define $\Delta y=\pm\sqrt{d^2-\Delta x^2}$. Otherwise if $\xi\geq\alpha$ we attempt to transmute a randomly selected solute or solvent. One Monte Carlo step consists of $L^2$ moves of any type and each move is accepted according to the usual Metropolis acceptance criterion.

We set the linear distance between two sites involved in a swap move to $d=4$ with the nearest integer approximation. The non-local swap move of impurities is introduced for fast equilibration of the system. The impurities can leave or enter the cluster efficiently with implementation of these moves. We note that this non-local impurity swap dynamics is different to the local (nearest neighbour) impurity swap dynamics studied in Ref.~\cite{mandal2021sm},  and so absolute estimated rates for the same $\alpha$ should not be compared directly.

For constant $d$ the parameter $\alpha$ controls the mean squared displacement of the mobile impurities with time, which increases monotonically with $\alpha$. We do not tune this parameter to match any particular real system, rather we explore the two limiting cases of static and "fast" impurities. The latter implies the spatial distribution of impurities equilibrates on timescales which are rapid compared to changes of cluster size. This could represent any system in which cluster growth is attachment limited and mobile impurities are present. 

We analyze the nucleation behaviour of the system by studying the nucleation free energy barrier for different impurity interaction energies. Since the formation of a post-critical cluster from the metastable phase at low temperatures is a rare event, unbiased simulations fail to sample the configuration space sufficiently within a tractable computational timescale. Therefore we use the umbrella sampling~\cite{us_torrie_1977,frenkel_2004_spherical_colloids} method to compute the nucleation free energy barrier as a function of solute cluster size. We use the standard geometric definition of a cluster, i.e. a set of solute particles that are contiguously connected via nearest neighbours to other solute particles.  An infinite square-well potential spanning a cluster size range of $20$ is used to bias the system to remain in a particular cluster size window, and we simulate multiple overlapping windows to cover the full range of relevant cluster sizes. The segments of the free energy curve obtained within each window are  then combined with the appropriate shift to reconstruct a smooth and continuous free energy curve. For windows which span smaller cluster sizes it is possible for multiple clusters to appear simultaneously that satisfy the size criterion of that window. It is important to count all such clusters and not just the largest in order to construct a free energy barrier consistent with CNT and the Becker-Doring-Zeldovich nucleation rate calculation. Otherwise the free energy exhibits a spurious minimum at the most frequently occurring largest cluster size observed in the metastable parent phase, which is grid size dependent. 

The nucleation rate,  i.e. the rate per unit area of forming post-critical nuclei from the metastable solution, is another important quantity that can be calculated either by direct simulation, or by using classical nucleation theory. Forward flux sampling~\cite{2009_ffs_allen,2009_tps_escobedo,2005_ffs_allen,polymer_folding_allen_2012} is a direct simulation method used for calculating the nucleation rate. In the forward flux sampling method, we define a set of ``interfaces'' at increasing values of the largest cluster size $\lambda$. The nucleation rate from the metastable solution phase can be written as
\be
\label{eq:ffs}
I=I_0\prod_{i=0}^{n}P(\lambda_{i+1}|\lambda_i),
\ee
where $I_0$ is the positive flux (crossings per unit time) measured through the zeroth interface in an unbiased simulation of the metastable phase. The quantity $P(\lambda_{i+1}|\lambda_i)$ is the probability of a simulation initialised at the $i$-th interface reaching the ($i$+1)-th interface before returning to the metastable solution. The interface $\lambda_0$ is chosen to be some small cluster size such that sufficient sampling of crossings can be sampled within a tractable simulation time. Use of the largest cluster size does result in a value of $I_0$ which is system size dependent for a given choice of $\lambda_0$, however the overall rate is not~\cite{blow_jcp_2023}.

The details of our implementation of both umbrella sampling and forward flux sampling algorithms could be found in Ref.~\cite{mandal2021sm}.

\section{Static impurities}
\label{sec:static}
\begin{figure}[t!]
\includegraphics[width=\columnwidth]{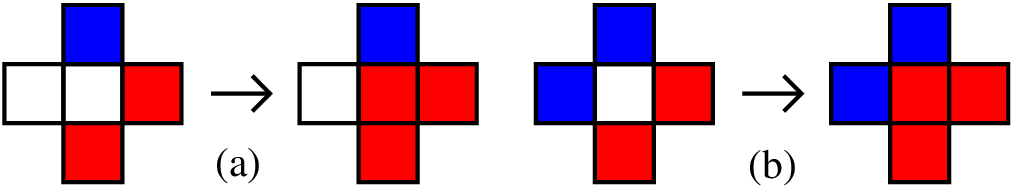}
\caption{The energy required to transmute a solvent (white) to a solute (red) surrounded by one and two impurities (blue) for the configurations depicted in (a) and (b) are respectively $\delta \epsilon=\epsilon_d-2J$ and $\delta \epsilon=2\epsilon_d-4J$. It  depends only on $\epsilon_d=\epsilon_+-\epsilon_-$ and $J$. The configurations before and after updating the central site are shown in left and right side of the arrow.}
\label{fig:delta_e}
\end{figure}

We wish to explore the full range of behaviour as a function of $\epsilon_+$ and $\epsilon_-$. 
In the case of static impurities only transmutation moves are performed. The interaction energies $\epsilon_{+}$ and $\epsilon_{-}$ appear in the acceptance probability for these moves via their difference $\epsilon_d=\epsilon_+-\epsilon_-$ since the moves replace solute-impurity interactions with an equal number of solvent-impurity interactions or vice versa. In Fig.~\ref{fig:delta_e} we have pictorially demonstrated the energy required in transmuting a solvent to a solute with one and two surrounded impurities. For static impurities is hence possible to reduce the exploration of interaction energies to a single parameter $\epsilon_d$. For convenience of comparison with later results, we choose to set $\epsilon_+=+\epsilon$ and $\epsilon_- = - \epsilon$ such that $\epsilon_d = 2\epsilon$. This sets impurity-solvent and impurity-solute interactions to be anti-symmetric in our simulations, however results for static impurities would be numerically identical for any choice of $\epsilon_+$ and $\epsilon_-$ that preserves $\epsilon_d$.

\begin{figure}[t!]
\includegraphics[width=\columnwidth]{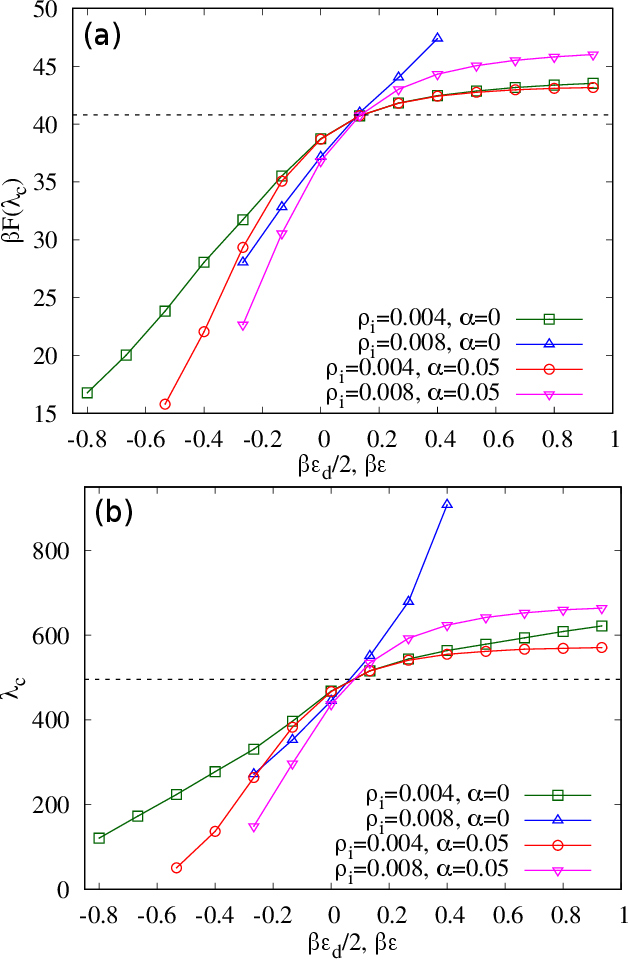}
\caption{Free energy barrier height (a) $F(\lambda_c)$ and (b) critical cluster size $\lambda_c$ as a function of the dimensionless magnitude of impurity interaction energy $\beta \epsilon$ for $\beta J=0.67$ and $\beta \Delta\mu=0.067$. Static ($\alpha=0$) and dynamic ($\alpha=0.05$) impurity cases with densities $\rho_i=0.004$ and $0.008$ are plotted. In both cases anti-symmetric impurity interactions ($\epsilon_- = -\epsilon_+)$ are used. The free energy barrier height tends toward saturation with increasing $\beta \epsilon$ for $\rho_i=0.004$ in the case of both static ($\alpha=0$) and dynamic ($\alpha=0.05$) impurities denoted by green squares and red circles respectively. At higher impurity density ($\rho_i=0.008$), the saturation is also observed in the case of dynamic impurities as denoted by magenta down-pointing triangles, but not in the case of static impurities as denoted by blue up-pointing triangles. Similar behaviour is also observed for $\lambda_c$ as a function of $\beta\epsilon$.
Dashed horizontal lines represent the barrier height $\beta F(\lambda_c)=40.81$ and critical cluster size $\lambda_c=496$ for the pure system ($\rho_i=0$) at $\beta J=0.67$ and $\beta \Delta\mu=0.067$.}
\label{fig:fc_lambdac}
\end{figure}

We calculate  the nucleation free energy $\beta F(\lambda)$ as a function of cluster size $\lambda$ over a range of the dimensionless impurity interaction strength $\beta \epsilon$ and at several values of static impurity density. Here $\beta=1/{k_BT}$ is the inverse temperature $T$, and $k_B$ is the Boltzmann constant which we set to $1$. In the free energy calculation, configurations in each umbrella sampling window are sampled over 48 realisations of the static impurity disorder and combined to compute a single free energy profile.
We then numerically estimate the critical cluster size $\lambda_c$ and free energy barrier height $\beta F(\lambda_c)$ for different values of $\beta \epsilon_d/2$. 
The free energy obtained from the umbrella sampling calculations is fitted by the free energy expression~\cite{cai2010pre,mandal2021sm} 
\be
\label{eq:mbd}
\beta F(\lambda)=-A_b\lambda+A_s\sqrt{\lambda}+\frac{5}{4}\log{\lambda}+B,
\ee
where $B=-\log{\rho_1}-A_s+A_b$, and $\rho_1$ is the density of isolated solute particles in the solution phase. The bulk term $A_b=\beta\Delta g$, where $\Delta g$ is the bulk free energy difference per particle between solute particles in the stable nucleating phase and in the metastable solution phase. At low temperatures, where the two phases are dominated by solute and solvent respectively, $\Delta g\approx \Delta \mu$. 
Estimation of $\lambda_c$ and $\beta F(\lambda_c)$ is done by fitting $\beta F(\lambda)$ to the expression given in Eq.~\ref{eq:mbd} using $A_b$ and $A_s$ as fitting parameters and finding the position and value of its maxima. However, if the free energy is not well fitted by Eq.~\ref{eq:mbd}, we add a higher order polynomial term, i.e., $\lambda^{3/2}$ with its prefactor as fitting parameter to improve estimation of $\lambda_c$. The higher order term is required for static impurities when $\beta \epsilon_d$ is large and positive. Here the free energy shows significant deviation from the expression shown in Eq.~\ref{eq:mbd} due to the strong repulsion and confinement effect from the impurities.  

The values of $\beta J$ we study are chosen to be greater than the critical inverse temperature of the two dimensional Ising model $\beta J_c=\ln(1+\sqrt{2})/2=0.4406\dots$~\cite{onsager_1944}. However, at high values of $\beta J$, when the corresponding temperature is low, the rare-event sampling algorithms used in this paper become inefficient. We hence focus on intermediate values of $\beta J$ which are $0.67$ and $0.83$, while doing umbrella sampling and forward flux sampling simulations. We note that performing simulations at different values of $\beta J$ implies performing it at different temperatures as the coupling constant is fixed to $J=1$. We choose different values of $\beta J$ in simulations to show the validity of our results for a range of temperature.

Variation of the free energy barrier height and the critical cluster size as a function of interaction energy difference $\beta \epsilon_d$, for $\rho_i=0.004$ (green squares) and $0.008$ (blue up-pointing triangles) at $\beta J=0.67$ and $\beta\Delta\mu=0.067$, with static impurities are shown in Fig.~\ref{fig:fc_lambdac}(a) and (b) respectively. The barrier height as well as critical cluster size increase with increasing $\beta \epsilon_d$.

This trend in free energy barrier is expected since the energy required to transmute a solvent to a solute (in the presence of a neighbouring impurity) is lowered when $\beta \epsilon_d$ is negative, meaning that nucleation can proceed preferentially in regions where impurities are present. For large negative $\beta \epsilon_d$, we would expect to see spinodal decomposition as a result of strong attraction between the impurities and solute.

For positive $\beta \epsilon_d$ nucleation will preferentially occur away from impurities and so the nucleation rate will saturate with respect to $\beta \epsilon_d$ provided there is sufficient space for a critical nucleus to form without needing to neighbour any impurity sites. This seems to be the case for $\rho_i = 0.004$, but not for the higher impurity density of $\rho_i=0.008$ where the barrier height continues to increase with $\beta \epsilon_d$ indicating that critical nuclei cannot form without encountering impurities that impede their formation.  

In Fig.~\ref{fig:fc_lambdac}(b), critical cluster size vs. $\beta \epsilon_d$ plots show similar behaviour, however growth is faster compared to the free energy barrier. See Fig. S3 in the Supporting Information (SI) for detailed free energy plots.

The barrier heights and critical sizes are consistent with our previous study when $\epsilon_d=0$~\cite{mandal2021sm}. 

We note that our free energy curves are computed by sampling configurations in each $\lambda$ window over several realisations of static disorder. Calculations at large and positive  $\beta \epsilon$ may be dominated by a small number of these realisations where sufficient room is available to form a critical nucleus without encountering impurities. Such clusters would have low energy and hence higher probability of formation when compared to equal size clusters in other realisations of the impurity disorder. This observation is related to that made in a recent study of the 3D random field Ising model~\cite{yao_heterogeneous} in which a spatially dependent reaction coordinate is used to account for the position, as well as the size of a nuclei, since the random field can create preferential locations for nuclei to form.

\section{Dynamic impurities}
\label{sec:dynamic}
In the case of dynamic impurities we include non-local impurity swap moves as described in section~\ref{sec:model}. The change in potential energy $\delta \Phi$ resulting from these moves depends on $J$, $\epsilon_+$ and $\epsilon_-$ explicitly and cannot be reduced to fewer parameters as in the static case. This is because impurities can hop between two sites that have differing numbers of solute and solvent neighbours. We study the system for different interaction types, e.g., symmetric ($\epsilon_- = \epsilon_+$), anti-symmetric ($\epsilon_- = -\epsilon_+$) and asymmetric ($|\epsilon_-| \neq |\epsilon_+|)$ with respect to impurity interactions with solute and solvent particles. We define the mobility parameter $\alpha$, which can take values between $0\leq\alpha <1$. The case $\alpha=0$ represents static impurities and $\alpha\neq0$ represents dynamic impurities with mobility increasing with $\alpha$. 
For the non-local impurity swap moves used in this study, equilibration of the impurity distribution is fast compared to the timescale on which clusters of solute particles grow or shrink. The free energy curves presented below hence represent a thermodynamically controlled nucleation process in which the impurity the distribution is sampled from a quasi-equilibrium distribution at each cluster size. 

\begin{figure}[t!]
\includegraphics[width=\columnwidth]{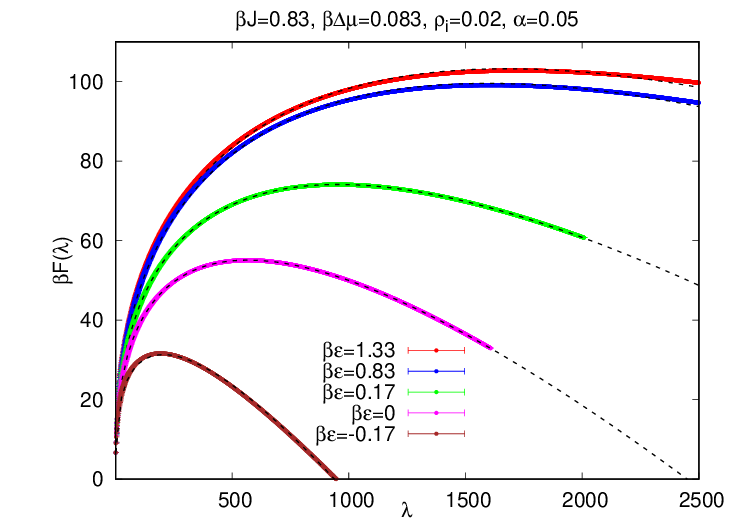}
\caption{Nucleation free energy with dynamic impurities, varying anti-symmetric interaction energy $\beta \epsilon_+=-\beta \epsilon_-=\beta \epsilon$ for $\beta J=0.83$, $\beta \Delta \mu=0.083$, $\rho_i=0.02$  and $\alpha=0.05$. Dotted lines are obtained from numerical fit of Eq.~\ref{eq:mbd} varying bulk $A_b$ and surface $A_s$ terms.}
\label{fig:free_sym_dyn-2}
\end{figure}

\subsection{Anti-symmetric interaction energy}
\label{subsec:anti-symmetric}
Here we set the interaction energies $\epsilon_+=-\epsilon_-=\epsilon$. Positive values of $\epsilon$ make impurity-solute interactions unfavourable and impurity-solvent interactions favourable. For negative values of $\epsilon$ the preference is reversed. A sequence of free energy curves as a function of cluster size $\lambda$ for different interaction energies $\beta \epsilon$ with $\beta J=0.83$, $\beta \Delta \mu=0.083$, $\rho_i=0.02$ and $\alpha=0.05$ is shown in Fig.~\ref{fig:free_sym_dyn-2}. The impurities act as nucleating sites for $\beta \epsilon< 0$ (lowering free energy barrier height), and repel solute for $\beta \epsilon>0$ (higher free energy barrier height). 

The trend in barrier height and critical cluster size with $\beta\epsilon$ is compared to the static impurity case in Fig.~\ref{fig:fc_lambdac}. For large positive values of $\beta \epsilon$ we expect the formation of clusters to exclude impurities even at high impurity density due to migration of impurities away from growing clusters (not possible for the static impurities). At impurity density $\rho_i=0.004$ and $\beta \epsilon\gtrsim0.53$, the impurity interaction energy with solute particles is sufficiently unfavourable that \emph{all} impurities are excluded from the growing cluster and there is no further change in barrier height as shown in Fig.~\ref{fig:fc_lambdac}(a) by red circles. The saturation threshold is $\beta \epsilon\gtrsim 0.8$ when $\rho_i$ is increased to $0.008$, denoted by magenta down-pointing triangles. Similar saturation is seen in critical cluster size as shown in Fig.~\ref{fig:fc_lambdac}(b). See Fig.~S1(a) in the SI for complete free energy plots.

A similar saturation in free energy barrier is observed even at impurity density $\rho_i=0.02$ and a different $\beta J=0.83$, but now at a more positive $\beta \epsilon\gtrsim1.17$. See Fig.~S1(b) in the SI for free energy plots. 

Unlike the static impurity case, this saturation of the barrier height with increasingly positive $\beta \epsilon$ results in a lower bound on the nucleation rate when the impurities are dynamic as the nucleation barrier height cannot increase further. As with the static case, there is no equivalent upper bound on the rate with dynamic impurities, as decreasingly negative $\beta \epsilon$  will lower the free energy of any clusters containing impurities until the spinodal limit is reached and the system can spontaneously transform into the solute-rich phase.

As the interaction strength $\beta \epsilon$ is made increasingly positive, we see a crossing both in $\beta F(\lambda_c)$ and $\lambda_c$ when these quantities are plotted for two different impurity densities, $\rho_i=0.004$ and $0.008$. This occurs at the value $\beta \epsilon^*\approx0.13$ for $\beta J=0.67$, $\beta \Delta\mu=0.067$ and $\alpha=0.05$ (see Fig.~\ref{fig:fc_lambdac}). Notably this is the value of $\beta \epsilon$ for which the system exhibits near identical barrier height and critical cluster size to the case where no impurities are present, implying cancellation of competing effects.

The role of impurities is opposite either side of this crossover. In region $\beta\epsilon>\beta\epsilon^*$, when impurities repel solute particles weakly, the barrier height is increased with increasing $\rho_i$. This is analogous to an observation in Ref.~\cite{Xu2024_additive}, where the growth of succinic acid is inhibited by different additives (glutaric acid, heptanedioic acid and azelaic acid), effectively increasing the interfacial tension with increasing additive concentration. For $\beta\epsilon<\beta\epsilon^*$, when there is a weak interaction between impurities and solute, interfacial tension decreases with increasing $\rho_i$ enhancing the nucleation rate. This is analogous to the experimental observation in Ref.~\cite{2018_surfactant}, where the presence of type-III antifreeze protein enhance the growth of ice nucleation by sitting at the boundaries of the cluster. A similar behaviour is obtained in a simulation of Potts lattice-gas in the presence of low dosage additives~\cite{peters_additive_2015}. The two curves cross where these competing effects cancel.

In Fig.~\ref{fig:fc_lambdac}, while comparing the dynamic ($\alpha=0.05$) vs. static ($\alpha=0$) plots, the barrier height and critical cluster size for dynamic impurities is always less than the values corresponding to static impurities, independent of interaction energy $\beta\epsilon$. This implies that the faster nucleation rate compared in the static case does not depend on the repulsive or attractive nature of the microscopic interactions when impurities are mobile. This behaviour is also observed for neutral impurity interactions in Ref.~\cite{mandal2021sm}.

\subsection{Symmetric interaction energy}
\label{subsec:symmetric}
Now we set the interaction energy $\epsilon_+=\epsilon_-=\epsilon$ to be symmetric. In this case the total solute-impurity and solvent-impurity interaction energy  takes the constant value  $4\epsilon\rho_iL^2$, independent of the impurity location, provided the impurities are at sufficiently low density that impurity-impurity interactions are negligible, which we expect to be the case for negative $\beta \epsilon$. The quantity $\rho_iL^2$ is the total number of impurity particles present in the solution. This impurity interaction term in the total energy can hence be treated as constant energy shift in the total energy which makes no difference to the nucleation behaviour. Thus we expect the free energy as well as the nucleation rate to be unchanged with respect to varying symmetric interaction energy $\epsilon$ as long as the impurities are sparsely distributed. 
In Fig.~S5 of the SI, we have plotted the variation of free energy for different negative values of symmetric interaction energy between impurity-solute and impurity-solvent, for $\beta J=0.83$, $\beta \Delta \mu=0.083$ and $\alpha=0.05$. We do not see a significant variation in free energy barrier for the range of (negative) $\beta \epsilon$ plotted in Fig.~S5 suggesting no impurity clustering occurs over that range. For positive $\beta \epsilon$ however, clustering of impurities will be expected, reducing the number of unfavourable impurity-solvent and impurity-solute interactions. This will be explored more generally in the next section.

\subsection{Behaviour map for asymmetric interactions}
\label{subsec:asymmetric}

\begin{figure}[t!]
\includegraphics[width=\columnwidth]{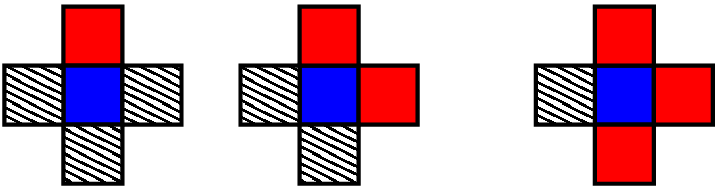}
\caption{Three types of impurity micro-states that contribute to the accumulation of impurities at the boundary of the nucleus. Red and blue boxes represent impurity and solute respectively. A shaded box could be either solvent or impurity.}
\label{fig:micro}
\end{figure}
\begin{figure*}[t!]
\includegraphics[width=2\columnwidth]{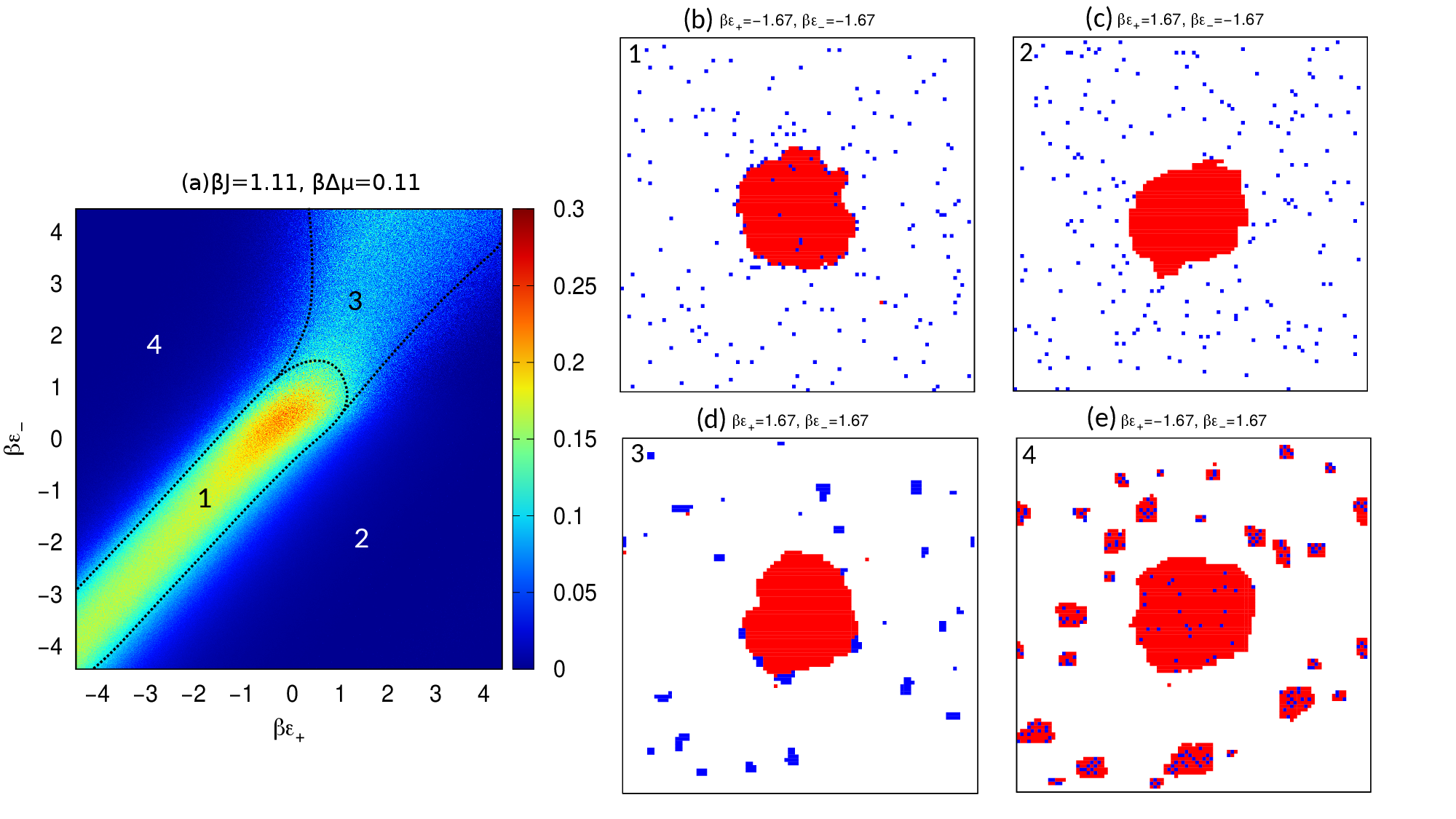}
\caption{(a) Variation of average impurity density at the boundary of the largest cluster $\phi$ as a function of $\beta\epsilon_+$ and $\beta\epsilon_-$ with dynamic impurities at density $\rho_i=0.02$, $\beta J=1.11$, $\beta \Delta\mu=0.11$ and $\alpha=0.05$. Depending on the positional occupancy of the impurities the interaction energy space could be divided into four regimes (1) surfactant: Impurities prefer to occupy the boundary of the cluster, (2) inert spectator: Impurities are completely excluded from the cluster without taking part in the nucleation process, (3) insoluble heterogeneous nucleation sites: Impurities form clusters which act as nucleation sites and (4) bulk stabilizer:  Impurities are completely inside the clusters stabilizing the bulk phase.  Snapshots of the system with biased simulations at quasi-equilibrium with largest cluster size $\lambda$ bounded between [800, 1000] at regime (b) 1, (c) 2, (d) 3 and (e) 4 of the behaviour map.}
\label{fig:phase_diagram_snap}
\end{figure*}

\begin{figure}[t!]
\includegraphics[width=\columnwidth]{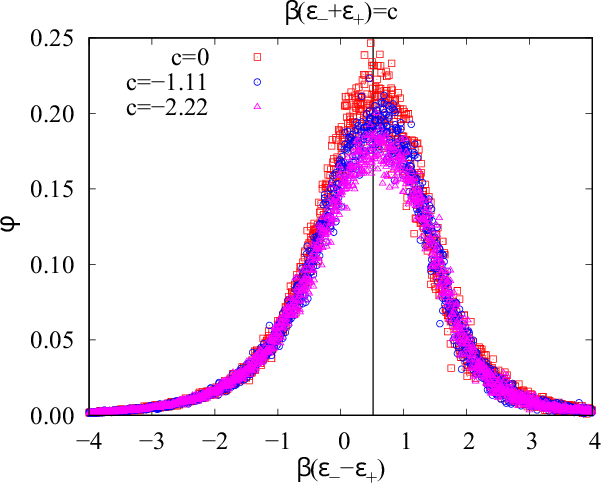}
\caption{One dimensional projection of the average impurity density at the boundary of the largest cluster, $\phi$ along $\beta(\epsilon_-+\epsilon_+)=c$ line for different $c$ at fixed impurities of density $\rho_i=0.02$ with $\beta J=1.11$ and $\beta\Delta\mu=0.11$. The maxima occurs at $\beta (\epsilon_--\epsilon_+)\approx 0.51$. The position of maxima is shown by vertical solid line.}
\label{fig:phase_diagram_proj}
\end{figure}

If allowing the impurity-solute and impurity-solvent interaction energies to be neither symmetric nor anti-symmetric, the available nucleation behaviour is richer.  We aim to map the possibilities by characterising the nucleation behaviour at each interaction choice. It is expected that for different interaction energies $\epsilon_+$ and $\epsilon_-$ the impurities preferentially occupy different positions in relation to the growing solute nucleus. These include impurities completely inside the clusters, completely outside the clusters and at the boundary of the solute clusters. To construct a map of this behaviour we calculate the average fraction of impurities that are located at the boundary of the largest cluster at fixed $\beta \epsilon_+$, $\beta \epsilon_-$ and $\beta \Delta \mu$. The local micro-state of impurities can be divided into five different groups depending on the number of nearest neighbour solutes (0 to 4). We count the fraction of impurities $\phi$ that have one, two or three nearest-neighbour solute particles as shown in Fig.~\ref{fig:micro}. This count takes place within a biased simulation in which we restrict the size of the largest cluster to be between $800$ to $1000$, and the size of the next largest cluster to be less than $30\%$ of the largest cluster size to avoid contacting with the largest cluster when the nucleation rate is high. We count only the impurity sites attached to the largest cluster and plot $\phi$ as a function of $\beta \epsilon_+$ and $\beta \epsilon_-$. The resulting map for $\beta J=1.11$, $\beta \Delta\mu=0.11$, $\rho_i=0.02$ and $\alpha=0.05$ is shown in Fig.~\ref{fig:phase_diagram_snap}(a). We have used a slightly higher value of $\beta J$ here as the preferential occupancy of impurities at the boundary of the growing cluster is only observed at low temperatures or equivalently at high $\beta J$~\cite{mandal2021sm}. With decreasing $\beta J$ the intensity of bright area decreases, making it difficult to differentiate between four regimes of the behaviour map introduced in next paragraph.

The behaviour map can be divided into four regimes depending on the positional occupancy and role of impurities. These regimes are (1) {\it surfactant}: The bright area where impurities prefer to occupy the boundary positions of a cluster acting as a surfactant, (2) {\it inert spectator}: The blue area in the right side of the behaviour map, where impurities are excluded from the nucleating clusters of solute, (3) {\it insoluble heterogeneous nucleating sites}: the noisy bright area at the top-right corner, where the impurities themselves form clusters which can act as heterogeneous nucleation sites (see section $\ref{sec:imp_clustering}$), and finally (4) {\it bulk stabilizer}: The blue area in the left side of the behaviour map, where impurities are preferentially located inside the red clusters as inclusions, stabilizing the bulk phase. Approximate boundary lines between regimes are  drawn by black dotted lines in Fig.~\ref{fig:phase_diagram_snap}(a). The previously discussed symmetric and anti-symmetric cases correspond to behaviour along the two diagonals. Snapshots of configurations from each of the four regimes in quasi-equilibrium with largest cluster size $\lambda$ confined between [800-1000], are shown in Fig.~\ref{fig:phase_diagram_snap}(b-e) for different interaction energies (b) $\beta\epsilon_+=-1.67$, $\beta\epsilon_-=-1.67$ (surfactant), (c) $\beta\epsilon_+=1.67$, $\beta\epsilon_-=-1.67$ (inert-spectator), (d) $\beta\epsilon_+=1.67$, $\beta\epsilon_-=1.67$ (insoluble impurity clusters) and (e) $\beta\epsilon_+=-1.67$, $\beta\epsilon_-=1.67$ (bulk-stabilizer).

An ad-hoc way of estimating the extent of regime (1) (impurities as surfactants) is the following. Impurities prefer to be in the solution phase and solute phase when $\epsilon_+>\epsilon_-$ and $\epsilon_+<\epsilon_-$ respectively. Arguably, $\epsilon_+=\epsilon_-$ would correspond to the regime where impurities prefer to occupy the boundary of a cluster. The expected width of region (1) on either side of this diagonal would be obtained by analyzing the stability of the interface. If we consider a flat interface without impurities separating pure solute and pure solvent regions, the interface energy per unit length would be $J$. On the other hand, if we add one layer of impurities at the interface between solute and solvent phase, the interface energy per unit length would be $(\epsilon_++\epsilon_-)/2$. The interface with impurities would be stable if the condition $(\epsilon_++\epsilon_-)/2<J$ is satisfied. Combining these two criterion, we obtain $\beta\epsilon<\beta J$, where $\epsilon_+=\epsilon_-=\epsilon$. We see that the derived condition for the width of surfactant regime is approximately satisfied in Fig.~\ref{fig:phase_diagram_snap}(a).

We analyze the impact of impurities in the regime outside the surfactant area and relate these to the trends in free energy observed in earlier sections.  In the surfactant regime, both solute and solvent attract impurities and we see surface accumulation of impurities. However, the free energy barrier does not change significantly with varying symmetric interaction energy. In the inert-spectator regime, solute repels but solvent attracts the impurities and we see strong exclusion of impurities from the nucleus. In this regime the barrier height remains unchanged with increasing symmetric interaction since all impurities are excluded from the nucleus. The regime in which impurities act as heterogeneous nucleation sites, both solute and solvent are repelled by the impurities. This strong repulsion forces impurities to form clusters. In the bulk-stabilizer regime, solute attracts but solvent repels impurities making nucleation strongly favourable with the impurities acting as nucleants. Here we see the presence of multiple clusters with impurities as inclusions. We also see a low barrier height to nucleation in this regime. 

It is evident from Fig.~\ref{fig:phase_diagram_snap}(a), that the impurities preferentially occupy the boundary positions of the cluster when $\epsilon_+\sim \epsilon_-$. But there exists a small asymmetry between these two energies as the maxima of the bright regions in Fig.~\ref{fig:phase_diagram_snap}(a) does not go exactly through the diagonal, i.e., the $\beta\epsilon_+=\beta\epsilon_-$ line. To illustrate this asymmetry we change the co-ordinate system of our behaviour map from $(\beta\epsilon_+, \beta\epsilon_-)$ to $(\beta\epsilon_++\beta\epsilon_-, \beta\epsilon_+-\beta\epsilon_-)$. Projections of the transformed behaviour map along $\beta(\epsilon_++\epsilon_-)=c$ line for different constant values of $c$ are shown in Fig.~\ref{fig:phase_diagram_proj}. The maxima of the projection plot occurs at $\beta(\epsilon_--\epsilon_+)\approx 0.51$ for $\beta J=1.11$ and $\beta \Delta\mu=0.11$. This implies impurities should have a small energy bias towards the solute compared to the solvent to maximise surface accumulation when the largest cluster is of the particular size used to construct this map.
We expect the presence of a transition with decreasing $\beta J$ (equivalent to increasing temperature when  $J$ is fixed) to a situation where impurities would no longer act as a surfactant for any type of interaction energy. This was analyzed for neutral impurities in Ref.~\cite{mandal2021sm}.

\section{Nucleation rate \& Becker-Doring-Zeldovich analysis}
\label{sec:rate}
We use forward flux sampling (FFS) to calculate the nucleation rate, i.e., the rate at which post-critical solute clusters that reach macroscopic size are obtained from the initial metastable solution phase. The mathematical expression for the nucleation rate within FFS is given in Eq.~\ref{eq:ffs}. The right hand side of Eq.~\ref{eq:ffs} may be interpreted as the rate of obtaining a cluster of size $\lambda=\lambda_{n+1}$ at the $(n+1)$-th interface, from the solution phase. The lower limiting value at which the decreasing rate converges for $\lambda > \lambda_c$ is the nucleation rate. We denote $L\times L$ Monte Carlo moves as the unit of time, i.e. one Monte Carlo step. In the case of static impurities, the nucleation rate $I$ is measured by the crossings per unit Monte Carlo step per single site which is consistent with the definition use by other authors~\cite{cai2010pre,ettori_2023}. For dynamic impurities, we divide the number of crossings per unit Monte Carlo step per single site by (1-$\alpha$), i.e. time is only progressed by attempted transmutation moves between solute and solvent.  Impurity dynamics are considered fast on this timescale such that the time elapsed during the non-local swap moves can be neglected.

\begin{figure}[t!]
\includegraphics[width=\columnwidth]{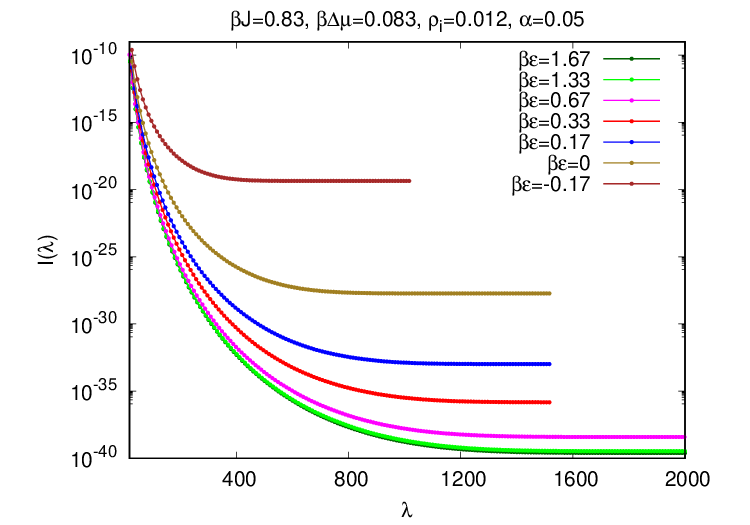}
\caption{Rate of obtaining a cluster of size $\lambda$ from the metastable solution phase for different anti-symmetric interaction energies $\beta \epsilon=\beta \epsilon_+=-\beta \epsilon_-$ at fixed $\beta J=0.83$, $\beta\Delta\mu=0.083$ and $\rho_i=0.012$ with dynamic ($\alpha=0.05$) impurities.}
\label{fig:rate_dyn}
\end{figure}

The rate $I(\lambda)$ of obtaining a cluster of size $\lambda$ starting from the solution phase is plotted in Fig.~\ref{fig:rate_dyn} for impurity density $\rho_i=0.012$ at $\beta J=0.83$, $\beta \Delta\mu=0.083$ and $\alpha=0.05$. Curves are plotted for a range of values of $\beta \epsilon$ that lie on the diagonal line joining the top-left (bulk stabilizer) to bottom-right (inert spectator) corner of the behaviour map  [see Fig.~\ref{fig:phase_diagram_snap}(a) for the behaviour map with impurity density $\rho_i=0.02$ at $\beta J=1.11$, $\beta \Delta\mu=0.11$ and $\alpha=0.05$]. The nucleation rate does not further decrease beyond a minimum value for $\beta \epsilon\gtrsim 1.33$ which belongs to the inert-spectator regime of the behaviour map. This represents the limit beyond which the probability of finding an impurity inside, or at the boundary of a large solute cluster is negligible and hence there is no further impact on the nucleation rate with increasing impurity interaction energy. 

\begin{figure}[t!]
\includegraphics[width=\columnwidth]{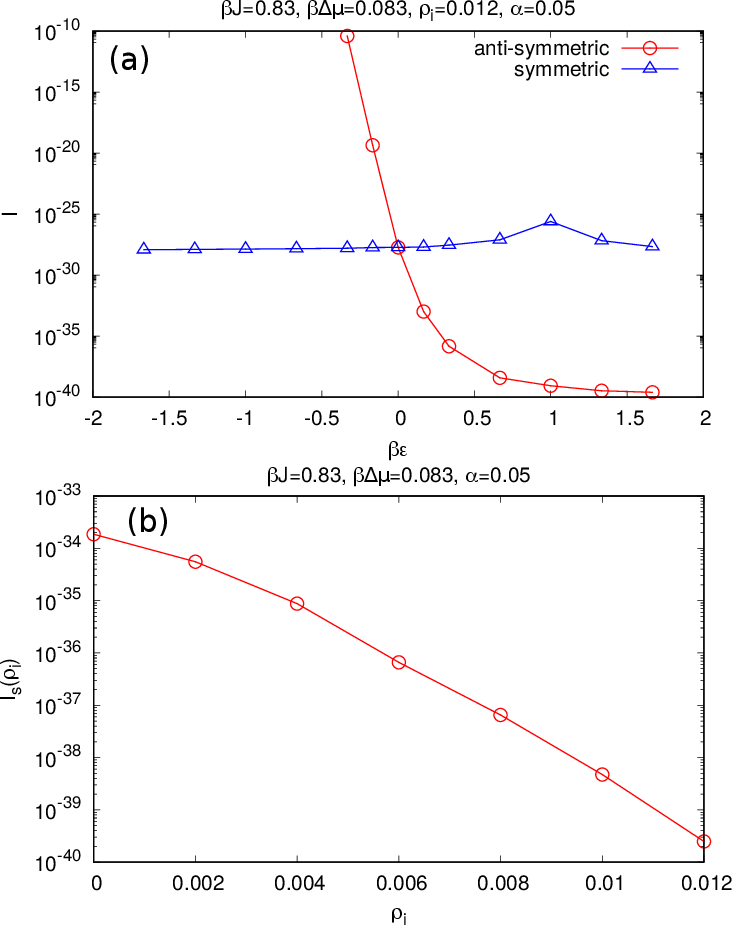}
\caption{(a) Nucleation rates as a function of interaction energy $\beta \epsilon$ for anti-symmetric and symmetric interaction energy at fixed $\beta J=0.83$, $\beta \Delta \mu=0.083$, $\rho_i=0.012$ and $\alpha=0.05$. (b) Saturated (with respect to increasing $\beta \epsilon$) nucleation rate $I_s(\rho_i)$ as a function of $\rho_i$ with anti-symmetric interaction energy with fixed $\beta J=0.83$, $\beta\Delta\mu=0.083$ and $\alpha=0.05$. The estimated standard error in calculating the nucleation rate plotted both in (a) and (b) is less than the size of the symbols.}
\label{fig:rate_diag}
\end{figure}

The nucleation rates extracted from Fig.~\ref{fig:rate_dyn} are re-plotted in Fig.~\ref{fig:rate_diag}(a) (red curve) and compared to the rates along the opposite diagonal of the behaviour map, where impurity interaction energies are symmetric (blue curve). Here the lower limit of the nucleation rate for anti-symmetric impurity interactions is visible as saturation of the rate $I$ with respect to increasingly positive $\beta \epsilon$. It can also be seen that the nucleation rate increases without apparent limit as $\beta \epsilon$ becomes more negative. This represents tending toward spinodal decomposition as the stability of the bulk solute phase is enhanced by the presence of impurity inclusions, similarly to the static case.

A natural question to ask is whether the minimum rate on increasing $\beta \epsilon$ for anti-symmetric interactions depends on impurity density. To answer that, we plot the saturated nucleation rate $I_s(\rho_i)$ for large and positive $\beta \epsilon$ and anti-symmetric interactions at different impurity density $\rho_i$ as shown in Fig.~\ref{fig:rate_diag}(b). We see that the saturated rate increases monotonically with decreasing $\rho_i$. As the impurities are excluded from the cluster all of them enter into the solution. The impurity density in the solution increases with increasing $\rho_i$. This excess impurities present in the solution change the effective chemical potential difference $\Delta\mu$ between the solution phase and the crystalline phase when $\rho_i$ is varied, changing the nucleation rate. We note that in the {\it inert spectator} regime while going from $\rho_i\neq 0$ (three-component solution) to $\rho_i= 0$ (two-component solution), the effective $\Delta \mu$ also changes which has a slight impact on nucleation rate.

In the case of symmetric impurity interaction energies, the two ends of the (blue) curve in Fig.~\ref{fig:rate_diag} lie in the {\it surfactant} regime (negative $\beta \epsilon$) and in the regime where impurities act as heterogeneous nucleation sites (positive $\beta \epsilon$). As expected from our analysis in subsection~\ref{subsec:symmetric}, we do not see a significant variation in the nucleation rate for symmetric interactions where $\beta \epsilon$ is negative and impurity-impurity interactions are rare. For large and positive $\beta\epsilon$ the limiting behaviour is that of a single large impurity cluster with a surface energy independent of whether it is surrounded by solute or solvent. Examination of simulation snapshots shows that the intermediate regime ($\beta\epsilon\approx 1)$ is characterised by the presence of both isolated impurities surrounded by solvent, and a single substantial impurity cluster. Nucleation of solute clusters occurs preferentially at the interface between this impurity cluster and the solution, but it is unclear why the nucleation rate is slightly enhanced in this case compared to larger values of $\beta\epsilon$ where \emph{all} impurities are present within a single cluster.

\textit{Becker-Doring-Zeldovich analysis:} Studying the ability of a Becker-Doring-Zeldovich analysis~\cite{cai2010pre,mandal2021sm} to reproduce trends in nucleation rate with varying impurity interactions can be instructive. In particularly it gives an understanding of which physical parameters (surface versus bulk free energies, kinetics) must be varied to fit the numerical simulation data and hence verify our mechanistic interpretation of results in the observed regimes.

We fit the free energies obtained from the umbrella sampling calculations with the modified free energy expression given in Eq.~\ref{eq:mbd}.
In the case of dynamic impurities, we use $A_b$ and $A_s$ as fitting parameters to fit Eq.(\ref{eq:mbd}) with the free energy curves obtained using umbrella sampling simulations, shown in Fig.~\ref{fig:free_sym_dyn-2} for $\rho_i=0.02$. The values of $A_b$ and $A_s$ obtained from the fitting is shown in Table~\ref{tab:bd}. These are used to calculate the nucleation rate described in the next paragraph. We see monotonic increase and decrease in $A_b$ and $A_s$ respectively with decreasing $\beta\epsilon$.

\begin{table}
\begin{center}
\begin{tabular}{|c c c|}
 \hline
 $\beta \epsilon$ & $A_b$ & $A_s$ \\ [0.5ex] 
 \hline\hline
 $1.33$ & $0.056$ & $4.54$ \\ [0.5ex] 
 \hline
 $0.83$ & $0.057$ & $4.47$   \\ [0.5ex] 
 \hline
 $0.17$ & $0.069$ & $4.16$  \\ [0.5ex] 
 \hline
  $0$ & $0.082$ & $3.81$  \\ [0.5ex] 
 \hline
  $-0.17$ & $0.114$ & $3.10$  \\ [0.5ex] 
 \hline
\end{tabular}
 \caption{Estimates of $A_b$ and $A_s$ from  fitting  Eq.(\ref{eq:mbd}) to the free energy plots obtained from umbrella sampling simulations shown in Fig.~\ref{fig:free_sym_dyn-2} for $\beta J=0.83$, $\beta \Delta\mu=0.083$, $\rho_i=0.02$ and $\alpha=0.05$ with anti-symmetric interaction energy and dynamic impurities with $\alpha=0.05$.}
  \label{tab:bd}
  \end{center}
\end{table}

The Becker-Doring-Zeldovich expression of the nucleation rate can be written as
\be
I_{BDZ}=D_c\Gamma e^{-\beta F(\lambda_c)},
\label{eq:rate_bd}
\ee
where $D_c$ is the diffusion coefficient, $\Gamma$ is the Zeldovich factor
\be
\Gamma=\sqrt{\frac{\beta}{2 \pi}\bigg[-\frac{\partial^2F(\lambda)}{\partial\lambda^2}\bigg|_{\lambda=\lambda_c}\bigg]},
\ee
evaluated at the critical cluster size $\lambda_c$. Using Eq.~\ref{eq:mbd}, these quantities can be expressed as
\bea
\Gamma&=&\sqrt{\frac{1}{2 \pi}\bigg(\frac{5}{4}{\lambda_c}^{-2}+\frac{1}{4}A_s{\lambda_c}^{-3/2}\bigg)},\\
\lambda_c&=&\bigg[\frac{A_s+\sqrt{A_s^2+20A_b}}{4A_b}\bigg]^2.
\eea
The diffusion coefficient $D_c=\langle {\Delta\lambda}(t)^2/2t \rangle$ is obtained from independent simulations,  starting from the critical cluster size and computing the slope of the mean square displacement versus time. See Fig.~S9 in SI for $D_c$ estimation.
In Table.~\ref{tab:one}, we compare the nucleation rates obtained from the Becker-Doring-Zeldovich analysis (see Eq.~\ref{eq:rate_bd}) and forward flux sampling simulation for different anti-symmetric interaction energies $\beta \epsilon$ at dynamic impurity density $\rho_i=0.004$.  The final two columns, $I_{BDZ}$ and $I_{FFS}$, are the rates obtained from Eq.~\ref{eq:rate_bd} and independent forward flux sampling simulations respectively. The maximum error in determining $I_{BDZ}$ is obtained by the expression 
\be
\frac{\Delta I_{BDZ}}{I_{BDZ}}=\frac{|\Delta D_c|}{D_c}+\frac{|\Delta \Gamma|}{\Gamma}+\beta |\Delta F(\lambda_c)|,
\ee
where $\Delta x$ is the error in determining the quantity $x$.
The excellent agreement between the results within the calculated error validates the application of CNT for the model studied, with only refitting of the surface and bulk terms due to the presence of impurities required. See Table.~S1 and Table.~S2 in the SI comparing $I_{BDZ}$ and $I_{FFS}$ at different regimes of the behaviour map.

\begin{table}
\begin{tabular}{|c c c c c c|}
 \hline
 $\beta \epsilon$ & $\lambda_c$ & $\beta F(\lambda_c)$ & $D_c$ & $I_{BDZ}$ & $I_{FFS}$ \\ [0.5ex] 
 \hline
 $0.8$ & $564$ & $42.94$ & $41.2$ & $2.8\times10^{-20}$ & $2.5\times10^{-20}$ \\ [0.5ex] 
 \hline
 $0.13$ & $510$ & $40.6$ & $39$ & $3.1\times10^{-19}$ & $3.1\times10^{-19}$  \\ [0.5ex] 
 \hline
 $0$ & $469$ & $38.8$ & $37.1$ & $1.8\times10^{-18}$ & $2.4\times10^{-18}$  \\ [0.5ex] 
 \hline
 $-0.13$ & $388$ & $35.2$ & $33.7$ & $6.7\times10^{-17}$ & $1.0\times10^{-16}$  \\ [0.5ex] 
 \hline
\end{tabular}
 \caption{Comparison of nucleation rates obtained from Becker-Doring-Zeldovich analysis ($I_{BDZ}$) and forward flux sampling  ($I_{FFS}$) for $\beta J=0.67$, $\beta \Delta \mu=0.067$,  $\rho_i=0.004$ and $\alpha=0.05$ with anti-symmetric interaction energy. The maximum error in determining $I_{BDZ}$ and $I_{FFS}$ are $72\%$ and $10\%$ respectively.}
 \label{tab:one}
\end{table}

\section{Impurity clustering and cross-nucleation} 
\label{sec:imp_clustering}
Impurities form multiple clusters when both impurity-solute and impurity-solvent interaction energies are positive, as seen in Fig.~\ref{fig:phase_diagram_snap}(d)]. Within this regime, when the repulsive interaction with solute and solvent is sufficiently strong, a single impurity cluster becomes the most stable configuration. We observe that nucleation of the solute phase starts from the boundaries of the impurity cluster, although the interaction between impurity and solute is repulsive.

\begin{figure}[t!]
\includegraphics[width=\columnwidth]{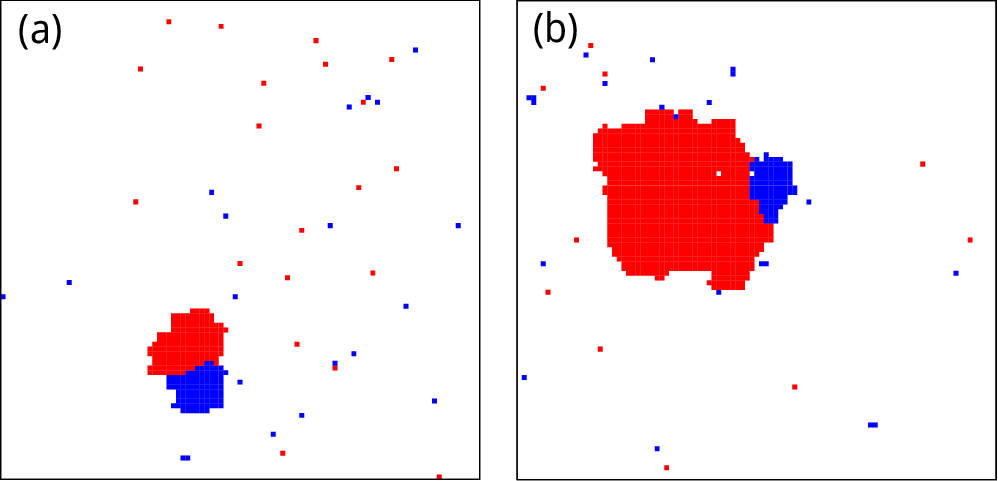}
\caption{Snapshots from the Umbrella sampling simulations, at (a) window-15 (solute cluster size lies between 150 and 170) and (b) window-104 (solute cluster size lies between 1040 and 1060),
showing the binding of an impurity clusters and solute cluster for symmetric interaction strength $\beta \epsilon=1.33$, $\beta J=0.83$, $\beta \Delta\mu=0.083$, $\rho_i=0.012$ and $\alpha=0.05$.}
\label{fig:imp_clust}
\end{figure}

Snapshots of a nucleating system, with symmetric interaction energy $\beta \epsilon=1.33$, $\beta J=0.83$, $\beta \Delta\mu=0.083$, $\rho_i=0.012$ and $\alpha=0.05$, obtained from the umbrella Sampling simulation for window-15 (largest cluster size lies between 150 and 170) and window-104 (largest cluster size lies between 1040 and 1060) are shown in Fig.~\ref{fig:imp_clust}(a) and (b) respectively. The binding between impurity and solute clusters occurs because the total surface energy of the two clusters is reduced when impurity-cluster and solute-cluster share a common boundary compared to the case when they are separated. 

Assuming a circular shape of both nuclei we may write the surface energy difference between the bonded-cluster and two separate-clusters as $\sigma_b(R,r)-\sigma_s(R,r)\approx -2rJ+\pi r(\epsilon_+-\epsilon_-)$ (see Appendix~\ref{appdx} for derivation), for $R\gg r$, where $R$ and $r$ are the radius of solute-cluster and impurity-cluster respectively. For symmetric interaction energy $\epsilon_+=\epsilon_-$, the second term in the right hand side vanishes and the surface energy difference becomes completely negative stabilising the bonded configuration. This is an example where the attraction between two clusters is induced by the microscopic repulsion between two particle types.  A similar idea has been used to calculate the free energy of a droplet starting to grow at the boundary of a parent nucleus~\cite{fletcher_1958}.

In the current context, this preferential formation of solute clusters at the boundary of impurity clusters can be considered as cross-nucleation~\cite{cross_nucleation1,cross_nucleation2}. Here, the impurity cluster acts as a heterogeneous nucleation site for the nucleation of solute clusters. 

\section{Conclusion}
\label{sec:conclusion}
We have studied the nucleation behaviour of a two dimensional Ising lattice-gas model in the presence of static and dynamic impurities with varying impurity-solute and impurity-solvent interaction energy.

In the case of static impurities, we have shown that the nucleation free energy barrier height
increases on increasing the difference between impurity-solute and impurity-solvent interaction energy $\epsilon_d$. The barrier height shows saturation with increasing $\epsilon_d$ when the static impurity density is low. However, we do not see such barrier height saturation when impurity density is high enough so that a critical cluster cannot fit into the largest void space between impurities. 

In the case of dynamic impurities, at high $\beta J$ (or equivalently, low temperatures) we observe preferential occupancy of the impurities at the boundary positions of the nucleus when the interaction energy of impurities with solute and solvent are similar. We have studied the system with varying the interaction energy and characterised four different nucleation regimes depending on the role and positional occupancy of impurities in the nucleation process. These regimes are {\it surfactant}, {\it inert spectator}, {\it heterogeneous nucleation sites} of impurity clusters and {\it bulk stabilizer}. Free energy behaviour and nucleation rate have been studied in each regime and the limits of impurity influence have been established. 
 
In this paper, we have the interactions between impurities to be neutral. Given the non-trivial behaviour when impurity clusters form, it would be interesting to extend this work for non-zero impurity-impurity interaction energy. How the different regimes in the behaviour map change with varying impurity-impurity interaction would also be interesting to investigate. It might be argued that the Monte Carlo moves in our current model limit the study to regimes where certain kinetic assumptions apply. It may be interesting to extend this to include diffusive transport of solute and solvent in place of transmutation moves, allow of concentration gradients of impurity, and other modifications to determine if new regimes of nucleation behaviour emerge. 

\section*{Author Contribution Statement}
DM developed the necessary simulation software, performed the simulations and wrote the manuscript. Both authors contributed to the design and interpretation of the simulations. DQ edited the manuscript.

\section*{Conflicts of Interest}
The authors have no conflicts of interest to declare.

\section*{Acknowledgement}
We acknowledge the support from the EPSRC Programme Grant (Grant EP/R018820/1) which funds the Crystallization in the Real World consortium. In addition, we gratefully acknowledge the use of the computational facilities provided by the University of Warwick Scientific Computing Research Technology Platform.

\section*{Data Availability}
Data associated with this manuscript is available via the University of Warwick Research Archive Portal at \url{https://wrap.warwick.ac.uk/187431/}.

\appendix
\section{Stability of bonded impurity-cluster and solute-cluster}
\label{appdx}
\begin{figure}[t!]
\centering
\includegraphics[width=0.4\columnwidth]{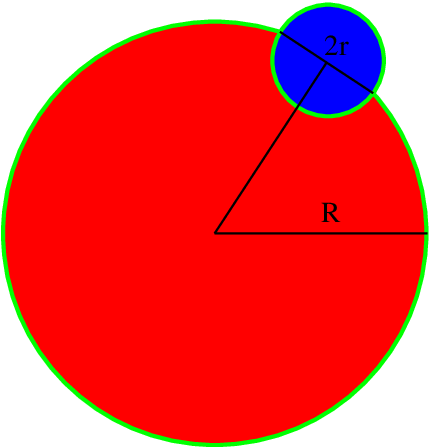}
\caption{A schematic diagram of cross nucleation to calculate the surface energy difference between the bonded and separate clusters.}
\label{fig:cross}
\end{figure}
A schematic representation of a bonded impurity cluster and solute cluster is shown in Fig.~\ref{fig:cross}. The surface energy of the bonded configuration $\sigma_b(R,r)$ can be written as the surface energy contribution obtained from the green boundary line which can be expressed as 
\be
\label{eq:sig_b}
\sigma_b(R,r)\approx(2\pi R-2r)J+\pi r(\epsilon_++\epsilon_-),
\ee
when we assume circular shape of the clusters. We also assume that radius of the solute-cluster does not change for the separate configuration which is true when $R\gg r$. Now, the total surface energy for separate configurations can be written as
\be
\label{eq:sig_s}
\sigma_s(R,r)\approx2\pi RJ+2\pi r\epsilon_-.
\ee
Subtracting Eq.~\ref{eq:sig_s} from Eq.~\ref{eq:sig_b} we find the surface energy difference $\sigma_b(R,r)-\sigma_s(R,r)\approx -2rJ+\pi r(\epsilon_+-\epsilon_-)$ which is independent of $R$. For symmetric impurity-solvent and impurity-solute interaction energy $\sigma_b(R,r)-\sigma_s(R,r)\approx -2rJ$. This implies that the bonded cluster has less surface energy compared to separate clusters. Similar analysis has been carried out to calculate the free energy of a droplet doing cross-nucleation~\cite{fletcher_1958}.


\begin{thebibliography}{38}%
\makeatletter
\providecommand \@ifxundefined [1]{%
 \@ifx{#1\undefined}
}%
\providecommand \@ifnum [1]{%
 \ifnum #1\expandafter \@firstoftwo
 \else \expandafter \@secondoftwo
 \fi
}%
\providecommand \@ifx [1]{%
 \ifx #1\expandafter \@firstoftwo
 \else \expandafter \@secondoftwo
 \fi
}%
\providecommand \natexlab [1]{#1}%
\providecommand \enquote  [1]{``#1''}%
\providecommand \bibnamefont  [1]{#1}%
\providecommand \bibfnamefont [1]{#1}%
\providecommand \citenamefont [1]{#1}%
\providecommand \href@noop [0]{\@secondoftwo}%
\providecommand \href [0]{\begingroup \@sanitize@url \@href}%
\providecommand \@href[1]{\@@startlink{#1}\@@href}%
\providecommand \@@href[1]{\endgroup#1\@@endlink}%
\providecommand \@sanitize@url [0]{\catcode `\\12\catcode `\$12\catcode
  `\&12\catcode `\#12\catcode `\^12\catcode `\_12\catcode `\%12\relax}%
\providecommand \@@startlink[1]{}%
\providecommand \@@endlink[0]{}%
\providecommand \url  [0]{\begingroup\@sanitize@url \@url }%
\providecommand \@url [1]{\endgroup\@href {#1}{\urlprefix }}%
\providecommand \urlprefix  [0]{URL }%
\providecommand \Eprint [0]{\href }%
\providecommand \doibase [0]{https://doi.org/}%
\providecommand \selectlanguage [0]{\@gobble}%
\providecommand \bibinfo  [0]{\@secondoftwo}%
\providecommand \bibfield  [0]{\@secondoftwo}%
\providecommand \translation [1]{[#1]}%
\providecommand \BibitemOpen [0]{}%
\providecommand \bibitemStop [0]{}%
\providecommand \bibitemNoStop [0]{.\EOS\space}%
\providecommand \EOS [0]{\spacefactor3000\relax}%
\providecommand \BibitemShut  [1]{\csname bibitem#1\endcsname}%
\let\auto@bib@innerbib\@empty
\bibitem [{\citenamefont {Becker}\ and\ \citenamefont
  {D{\"o}ring}(1935)}]{1935-becker-aphys}%
  \BibitemOpen
  \bibfield  {author} {\bibinfo {author} {\bibfnamefont {R.}~\bibnamefont
  {Becker}}\ and\ \bibinfo {author} {\bibfnamefont {W.}~\bibnamefont
  {D{\"o}ring}},\ }\bibfield  {title} {\enquote {\bibinfo {title} {The kinetic
  treatment of nuclear formation in supersaturated vapors},}\ }\href@noop {}
  {\bibfield  {journal} {\bibinfo  {journal} {Ann. Phys}\ }\textbf {\bibinfo
  {volume} {24}},\ \bibinfo {pages} {752} (\bibinfo {year} {1935})}\BibitemShut
  {NoStop}%
\bibitem [{\citenamefont {Sunyaev}(1992)}]{zeldovich}%
  \BibitemOpen
  \bibinfo {editor} {\bibfnamefont {R.~A.}\ \bibnamefont {Sunyaev}},\ ed.,\
  \enquote {\bibinfo {title} {10. on the theory of new phase formation.
  cavitation},}\ in\ \href@noop {} {\emph {\bibinfo {booktitle} {Selected Works
  of Yakov Borisovich Zeldovich, Volume I}}}\ (\bibinfo  {publisher} {Princeton
  University Press},\ \bibinfo {address} {Princeton},\ \bibinfo {year} {1992})\
  pp.\ \bibinfo {pages} {120--137}\BibitemShut {NoStop}%
\bibitem [{\citenamefont {Kashchiev}(2000)}]{Kashchiev2000}%
  \BibitemOpen
  \bibfield  {author} {\bibinfo {author} {\bibfnamefont {D.}~\bibnamefont
  {Kashchiev}},\ }\href@noop {} {\emph {\bibinfo {title} {Nucleation}}}\
  (\bibinfo  {publisher} {Butterworth-Heinemann},\ \bibinfo {address}
  {Oxford},\ \bibinfo {year} {2000})\BibitemShut {NoStop}%
\bibitem [{\citenamefont {Ryu}\ and\ \citenamefont {Cai}(2010)}]{cai2010pre}%
  \BibitemOpen
  \bibfield  {author} {\bibinfo {author} {\bibfnamefont {S.}~\bibnamefont
  {Ryu}}\ and\ \bibinfo {author} {\bibfnamefont {W.}~\bibnamefont {Cai}},\
  }\bibfield  {title} {\enquote {\bibinfo {title} {Numerical tests of
  nucleation theories for the ising models},}\ }\href@noop {} {\bibfield
  {journal} {\bibinfo  {journal} {Phys. Rev. E}\ }\textbf {\bibinfo {volume}
  {82}},\ \bibinfo {pages} {011603} (\bibinfo {year} {2010})}\BibitemShut
  {NoStop}%
\bibitem [{\citenamefont {Stauffer}, \citenamefont {Coniglio},\ and\
  \citenamefont {Heermann}(1982)}]{mc-prl-1982}%
  \BibitemOpen
  \bibfield  {author} {\bibinfo {author} {\bibfnamefont {D.}~\bibnamefont
  {Stauffer}}, \bibinfo {author} {\bibfnamefont {A.}~\bibnamefont {Coniglio}},\
  and\ \bibinfo {author} {\bibfnamefont {D.~W.}\ \bibnamefont {Heermann}},\
  }\bibfield  {title} {\enquote {\bibinfo {title} {Monte carlo experiment for
  nucleation rate in the three-dimensional ising model},}\ }\href@noop {}
  {\bibfield  {journal} {\bibinfo  {journal} {Phys. Rev. Lett.}\ }\textbf
  {\bibinfo {volume} {49}},\ \bibinfo {pages} {1299--1302} (\bibinfo {year}
  {1982})}\BibitemShut {NoStop}%
\bibitem [{\citenamefont {Shneidman}, \citenamefont {Jackson},\ and\
  \citenamefont {Beatty}(1999)}]{cnt-jcp-1999}%
  \BibitemOpen
  \bibfield  {author} {\bibinfo {author} {\bibfnamefont {V.~A.}\ \bibnamefont
  {Shneidman}}, \bibinfo {author} {\bibfnamefont {K.~A.}\ \bibnamefont
  {Jackson}},\ and\ \bibinfo {author} {\bibfnamefont {K.~M.}\ \bibnamefont
  {Beatty}},\ }\bibfield  {title} {\enquote {\bibinfo {title} {{On the
  applicability of the classical nucleation theory in an Ising system}},}\
  }\href@noop {} {\bibfield  {journal} {\bibinfo  {journal} {J. Chem. Phys.}\
  }\textbf {\bibinfo {volume} {111}},\ \bibinfo {pages} {6932--6941} (\bibinfo
  {year} {1999})}\BibitemShut {NoStop}%
\bibitem [{\citenamefont {Schmitz}, \citenamefont {Virnau},\ and\ \citenamefont
  {Binder}(2013)}]{mc-pre-2013-binder}%
  \BibitemOpen
  \bibfield  {author} {\bibinfo {author} {\bibfnamefont {F.}~\bibnamefont
  {Schmitz}}, \bibinfo {author} {\bibfnamefont {P.}~\bibnamefont {Virnau}},\
  and\ \bibinfo {author} {\bibfnamefont {K.}~\bibnamefont {Binder}},\
  }\bibfield  {title} {\enquote {\bibinfo {title} {Monte carlo tests of
  nucleation concepts in the lattice gas model},}\ }\href@noop {} {\bibfield
  {journal} {\bibinfo  {journal} {Phys. Rev. E}\ }\textbf {\bibinfo {volume}
  {87}},\ \bibinfo {pages} {053302} (\bibinfo {year} {2013})}\BibitemShut
  {NoStop}%
\bibitem [{\citenamefont {Duff}\ and\ \citenamefont
  {Peters}(2009)}]{duff_jcp_2009}%
  \BibitemOpen
  \bibfield  {author} {\bibinfo {author} {\bibfnamefont {N.}~\bibnamefont
  {Duff}}\ and\ \bibinfo {author} {\bibfnamefont {B.}~\bibnamefont {Peters}},\
  }\bibfield  {title} {\enquote {\bibinfo {title} {Nucleation in a potts
  lattice gas model of crystallization from solution},}\ }\href@noop {}
  {\bibfield  {journal} {\bibinfo  {journal} {J. Chem. Phys.}\ }\textbf
  {\bibinfo {volume} {131}},\ \bibinfo {pages} {184101} (\bibinfo {year}
  {2009})}\BibitemShut {NoStop}%
\bibitem [{\citenamefont {Lifanov}, \citenamefont {Vorselaars},\ and\
  \citenamefont {Quigley}(2016)}]{lifanov_jcp_2016}%
  \BibitemOpen
  \bibfield  {author} {\bibinfo {author} {\bibfnamefont {Y.}~\bibnamefont
  {Lifanov}}, \bibinfo {author} {\bibfnamefont {B.}~\bibnamefont
  {Vorselaars}},\ and\ \bibinfo {author} {\bibfnamefont {D.}~\bibnamefont
  {Quigley}},\ }\bibfield  {title} {\enquote {\bibinfo {title} {Nucleation
  barrier reconstruction via the seeding method in a lattice model with
  competing nucleation pathways},}\ }\href@noop {} {\bibfield  {journal}
  {\bibinfo  {journal} {J. Chem. Phys.}\ }\textbf {\bibinfo {volume} {145}},\
  \bibinfo {pages} {211912} (\bibinfo {year} {2016})}\BibitemShut {NoStop}%
\bibitem [{\citenamefont {Teychené}, \citenamefont {Rodríguez-Ruiz},\ and\
  \citenamefont {Ramamoorthy}(2020)}]{TEYCHENE20201}%
  \BibitemOpen
  \bibfield  {author} {\bibinfo {author} {\bibfnamefont {S.}~\bibnamefont
  {Teychené}}, \bibinfo {author} {\bibfnamefont {I.}~\bibnamefont
  {Rodríguez-Ruiz}},\ and\ \bibinfo {author} {\bibfnamefont {R.~K.}\
  \bibnamefont {Ramamoorthy}},\ }\bibfield  {title} {\enquote {\bibinfo {title}
  {Reactive crystallization: From mixing to control of kinetics by
  additives},}\ }\href@noop {} {\bibfield  {journal} {\bibinfo  {journal}
  {Current Opinion in Colloid \& Interface Science}\ }\textbf {\bibinfo
  {volume} {46}},\ \bibinfo {pages} {1--19} (\bibinfo {year} {2020})},\
  \bibinfo {note} {special Topic: Colloidal and Interfacial Challenges Related
  to Separations, Analysis and Recycling}\BibitemShut {NoStop}%
\bibitem [{\citenamefont {Vorontsov}\ \emph {et~al.}(2018)\citenamefont
  {Vorontsov}, \citenamefont {Sazaki}, \citenamefont {Titaeva}, \citenamefont
  {Kim}, \citenamefont {Bayer-Giraldi},\ and\ \citenamefont
  {Furukawa}}]{2018_surfactant}%
  \BibitemOpen
  \bibfield  {author} {\bibinfo {author} {\bibfnamefont {D.~A.}\ \bibnamefont
  {Vorontsov}}, \bibinfo {author} {\bibfnamefont {G.}~\bibnamefont {Sazaki}},
  \bibinfo {author} {\bibfnamefont {E.~K.}\ \bibnamefont {Titaeva}}, \bibinfo
  {author} {\bibfnamefont {E.~L.}\ \bibnamefont {Kim}}, \bibinfo {author}
  {\bibfnamefont {M.}~\bibnamefont {Bayer-Giraldi}},\ and\ \bibinfo {author}
  {\bibfnamefont {Y.}~\bibnamefont {Furukawa}},\ }\bibfield  {title} {\enquote
  {\bibinfo {title} {Growth of ice crystals in the presence of type iii
  antifreeze protein},}\ }\href@noop {} {\bibfield  {journal} {\bibinfo
  {journal} {Crystal Growth {\&} Design}\ }\textbf {\bibinfo {volume} {18}},\
  \bibinfo {pages} {2563--2571} (\bibinfo {year} {2018})}\BibitemShut {NoStop}%
\bibitem [{\citenamefont {Gebauer}\ \emph {et~al.}(2009)\citenamefont
  {Gebauer}, \citenamefont {Cölfen}, \citenamefont {Verch},\ and\
  \citenamefont {Antonietti}}]{additive_caco3_2009}%
  \BibitemOpen
  \bibfield  {author} {\bibinfo {author} {\bibfnamefont {D.}~\bibnamefont
  {Gebauer}}, \bibinfo {author} {\bibfnamefont {H.}~\bibnamefont {Cölfen}},
  \bibinfo {author} {\bibfnamefont {A.}~\bibnamefont {Verch}},\ and\ \bibinfo
  {author} {\bibfnamefont {M.}~\bibnamefont {Antonietti}},\ }\bibfield  {title}
  {\enquote {\bibinfo {title} {The multiple roles of additives in caco3
  crystallization: A quantitative case study},}\ }\href@noop {} {\bibfield
  {journal} {\bibinfo  {journal} {Advanced Materials}\ }\textbf {\bibinfo
  {volume} {21}},\ \bibinfo {pages} {435--439} (\bibinfo {year}
  {2009})}\BibitemShut {NoStop}%
\bibitem [{\citenamefont {Xu}\ \emph {et~al.}(2024)\citenamefont {Xu},
  \citenamefont {Liu}, \citenamefont {Zhang},\ and\ \citenamefont
  {Wang}}]{Xu2024_additive}%
  \BibitemOpen
  \bibfield  {author} {\bibinfo {author} {\bibfnamefont {S.}~\bibnamefont
  {Xu}}, \bibinfo {author} {\bibfnamefont {Y.}~\bibnamefont {Liu}}, \bibinfo
  {author} {\bibfnamefont {R.}~\bibnamefont {Zhang}},\ and\ \bibinfo {author}
  {\bibfnamefont {Y.}~\bibnamefont {Wang}},\ }\bibfield  {title} {\enquote
  {\bibinfo {title} {How structurally similar additives affect the
  dimensionless interfacial tension for inducing nucleation: The case study of
  succinic acid},}\ }\href@noop {} {\bibfield  {journal} {\bibinfo  {journal}
  {Crystal Growth {\&} Design}\ }\textbf {\bibinfo {volume} {24}},\ \bibinfo
  {pages} {103--114} (\bibinfo {year} {2024})}\BibitemShut {NoStop}%
\bibitem [{\citenamefont {Han}\ \emph {et~al.}(2019)\citenamefont {Han},
  \citenamefont {Wang}, \citenamefont {Yang}, \citenamefont {Gong},
  \citenamefont {Chen},\ and\ \citenamefont {Gong}}]{2019_additive_surface}%
  \BibitemOpen
  \bibfield  {author} {\bibinfo {author} {\bibfnamefont {D.}~\bibnamefont
  {Han}}, \bibinfo {author} {\bibfnamefont {Y.}~\bibnamefont {Wang}}, \bibinfo
  {author} {\bibfnamefont {Y.}~\bibnamefont {Yang}}, \bibinfo {author}
  {\bibfnamefont {T.}~\bibnamefont {Gong}}, \bibinfo {author} {\bibfnamefont
  {Y.}~\bibnamefont {Chen}},\ and\ \bibinfo {author} {\bibfnamefont
  {J.}~\bibnamefont {Gong}},\ }\bibfield  {title} {\enquote {\bibinfo {title}
  {Revealing the role of a surfactant in the nucleation and crystal growth of
  thiamine nitrate: experiments and simulation studies},}\ }\href@noop {}
  {\bibfield  {journal} {\bibinfo  {journal} {CrystEngComm}\ }\textbf {\bibinfo
  {volume} {21}},\ \bibinfo {pages} {3576--3585} (\bibinfo {year}
  {2019})}\BibitemShut {NoStop}%
\bibitem [{\citenamefont {Anwar}\ \emph {et~al.}(2009)\citenamefont {Anwar},
  \citenamefont {Boateng}, \citenamefont {Tamaki},\ and\ \citenamefont
  {Odedra}}]{anwar_2009}%
  \BibitemOpen
  \bibfield  {author} {\bibinfo {author} {\bibfnamefont {J.}~\bibnamefont
  {Anwar}}, \bibinfo {author} {\bibfnamefont {P.}~\bibnamefont {Boateng}},
  \bibinfo {author} {\bibfnamefont {R.}~\bibnamefont {Tamaki}},\ and\ \bibinfo
  {author} {\bibfnamefont {S.}~\bibnamefont {Odedra}},\ }\bibfield  {title}
  {\enquote {\bibinfo {title} {Mode of action and design rules for additives
  that modulate crystal nucleation},}\ }\href@noop {} {\bibfield  {journal}
  {\bibinfo  {journal} {Angewandte Chemie International Edition}\ }\textbf
  {\bibinfo {volume} {48}},\ \bibinfo {pages} {1596--1600} (\bibinfo {year}
  {2009})}\BibitemShut {NoStop}%
\bibitem [{\citenamefont {Anwar}\ and\ \citenamefont
  {Zahn}(2011)}]{anwar_2011}%
  \BibitemOpen
  \bibfield  {author} {\bibinfo {author} {\bibfnamefont {J.}~\bibnamefont
  {Anwar}}\ and\ \bibinfo {author} {\bibfnamefont {D.}~\bibnamefont {Zahn}},\
  }\bibfield  {title} {\enquote {\bibinfo {title} {Uncovering molecular
  processes in crystal nucleation and growth by using molecular simulation},}\
  }\href@noop {} {\bibfield  {journal} {\bibinfo  {journal} {Angewandte Chemie
  International Edition}\ }\textbf {\bibinfo {volume} {50}},\ \bibinfo {pages}
  {1996--2013} (\bibinfo {year} {2011})}\BibitemShut {NoStop}%
\bibitem [{\citenamefont {Bertolazzo}\ \emph {et~al.}(2018)\citenamefont
  {Bertolazzo}, \citenamefont {Naullage}, \citenamefont {Peters},\ and\
  \citenamefont {Molinero}}]{bertolazzo2018}%
  \BibitemOpen
  \bibfield  {author} {\bibinfo {author} {\bibfnamefont {A.~A.}\ \bibnamefont
  {Bertolazzo}}, \bibinfo {author} {\bibfnamefont {P.~M.}\ \bibnamefont
  {Naullage}}, \bibinfo {author} {\bibfnamefont {B.}~\bibnamefont {Peters}},\
  and\ \bibinfo {author} {\bibfnamefont {V.}~\bibnamefont {Molinero}},\
  }\bibfield  {title} {\enquote {\bibinfo {title} {The clathrate-water
  interface is oleophilic},}\ }\href@noop {} {\bibfield  {journal} {\bibinfo
  {journal} {The Journal of Physical Chemistry Letters}\ }\textbf {\bibinfo
  {volume} {9}},\ \bibinfo {pages} {3224--3231} (\bibinfo {year}
  {2018})}\BibitemShut {NoStop}%
\bibitem [{\citenamefont {Poon}\ \emph {et~al.}(2017)\citenamefont {Poon},
  \citenamefont {Lemke}, \citenamefont {Peter}, \citenamefont {Molinero},\ and\
  \citenamefont {Peters}}]{poon2017}%
  \BibitemOpen
  \bibfield  {author} {\bibinfo {author} {\bibfnamefont {G.~G.}\ \bibnamefont
  {Poon}}, \bibinfo {author} {\bibfnamefont {T.}~\bibnamefont {Lemke}},
  \bibinfo {author} {\bibfnamefont {C.}~\bibnamefont {Peter}}, \bibinfo
  {author} {\bibfnamefont {V.}~\bibnamefont {Molinero}},\ and\ \bibinfo
  {author} {\bibfnamefont {B.}~\bibnamefont {Peters}},\ }\bibfield  {title}
  {\enquote {\bibinfo {title} {Soluble oligomeric nucleants: Simulations of
  chain length, binding strength, and volume fraction effects.}}\ }\href@noop
  {} {\bibfield  {journal} {\bibinfo  {journal} {The journal of physical
  chemistry letters}\ }\textbf {\bibinfo {volume} {8 23}},\ \bibinfo {pages}
  {5815--5820} (\bibinfo {year} {2017})}\BibitemShut {NoStop}%
\bibitem [{\citenamefont {Hudait}\ and\ \citenamefont
  {Molinero}(2014)}]{hudait2014}%
  \BibitemOpen
  \bibfield  {author} {\bibinfo {author} {\bibfnamefont {A.}~\bibnamefont
  {Hudait}}\ and\ \bibinfo {author} {\bibfnamefont {V.}~\bibnamefont
  {Molinero}},\ }\bibfield  {title} {\enquote {\bibinfo {title} {Ice
  crystallization in ultrafine water--salt aerosols: Nucleation, ice-solution
  equilibrium, and internal structure},}\ }\href@noop {} {\bibfield  {journal}
  {\bibinfo  {journal} {Journal of the American Chemical Society}\ }\textbf
  {\bibinfo {volume} {136}},\ \bibinfo {pages} {8081--8093} (\bibinfo {year}
  {2014})}\BibitemShut {NoStop}%
\bibitem [{\citenamefont {Mandal}\ and\ \citenamefont
  {Quigley}(2021)}]{mandal2021sm}%
  \BibitemOpen
  \bibfield  {author} {\bibinfo {author} {\bibfnamefont {D.}~\bibnamefont
  {Mandal}}\ and\ \bibinfo {author} {\bibfnamefont {D.}~\bibnamefont
  {Quigley}},\ }\bibfield  {title} {\enquote {\bibinfo {title} {Nucleation rate
  in the two dimensional ising model in the presence of random impurities},}\
  }\href@noop {} {\bibfield  {journal} {\bibinfo  {journal} {Soft Matter}\
  }\textbf {\bibinfo {volume} {17}},\ \bibinfo {pages} {8642--8650} (\bibinfo
  {year} {2021})}\BibitemShut {NoStop}%
\bibitem [{\citenamefont {Page}\ and\ \citenamefont
  {Sear}(2006)}]{page2006prl}%
  \BibitemOpen
  \bibfield  {author} {\bibinfo {author} {\bibfnamefont {A.~J.}\ \bibnamefont
  {Page}}\ and\ \bibinfo {author} {\bibfnamefont {R.~P.}\ \bibnamefont
  {Sear}},\ }\bibfield  {title} {\enquote {\bibinfo {title} {Heterogeneous
  nucleation in and out of pores},}\ }\href@noop {} {\bibfield  {journal}
  {\bibinfo  {journal} {Phys. Rev. Lett.}\ }\textbf {\bibinfo {volume} {97}},\
  \bibinfo {pages} {065701} (\bibinfo {year} {2006})}\BibitemShut {NoStop}%
\bibitem [{\citenamefont {Hedges}\ and\ \citenamefont
  {Whitelam}(2012)}]{whitelam2012sm}%
  \BibitemOpen
  \bibfield  {author} {\bibinfo {author} {\bibfnamefont {L.~O.}\ \bibnamefont
  {Hedges}}\ and\ \bibinfo {author} {\bibfnamefont {S.}~\bibnamefont
  {Whitelam}},\ }\bibfield  {title} {\enquote {\bibinfo {title} {Patterning a
  surface so as to speed nucleation from solution},}\ }\href@noop {} {\bibfield
   {journal} {\bibinfo  {journal} {Soft Matt.}\ }\textbf {\bibinfo {volume}
  {8}},\ \bibinfo {pages} {8624--8635} (\bibinfo {year} {2012})}\BibitemShut
  {NoStop}%
\bibitem [{\citenamefont {Grosfils}\ and\ \citenamefont
  {Lutsko}(2021)}]{lutsko2021}%
  \BibitemOpen
  \bibfield  {author} {\bibinfo {author} {\bibfnamefont {P.}~\bibnamefont
  {Grosfils}}\ and\ \bibinfo {author} {\bibfnamefont {J.~F.}\ \bibnamefont
  {Lutsko}},\ }\bibfield  {title} {\enquote {\bibinfo {title} {Impact of
  surface roughness on crystal nucleation},}\ }\href@noop {} {\bibfield
  {journal} {\bibinfo  {journal} {Crystals}\ }\textbf {\bibinfo {volume}
  {11}},\ \bibinfo {pages} {4} (\bibinfo {year} {2021})}\BibitemShut {NoStop}%
\bibitem [{\citenamefont {Mandal}\ and\ \citenamefont
  {Quigley}(2022)}]{2022_mandal}%
  \BibitemOpen
  \bibfield  {author} {\bibinfo {author} {\bibfnamefont {D.}~\bibnamefont
  {Mandal}}\ and\ \bibinfo {author} {\bibfnamefont {D.}~\bibnamefont
  {Quigley}},\ }\bibfield  {title} {\enquote {\bibinfo {title} {{Kinetic
  control of competing nuclei in a dimer lattice-gas model}},}\ }\href@noop {}
  {\bibfield  {journal} {\bibinfo  {journal} {The Journal of Chemical Physics}\
  }\textbf {\bibinfo {volume} {157}},\ \bibinfo {pages} {214501} (\bibinfo
  {year} {2022})}\BibitemShut {NoStop}%
\bibitem [{\citenamefont {Ettori}, \citenamefont {Sluckin},\ and\ \citenamefont
  {Biscari}(2023)}]{ettori_2023}%
  \BibitemOpen
  \bibfield  {author} {\bibinfo {author} {\bibfnamefont {F.}~\bibnamefont
  {Ettori}}, \bibinfo {author} {\bibfnamefont {T.~J.}\ \bibnamefont
  {Sluckin}},\ and\ \bibinfo {author} {\bibfnamefont {P.}~\bibnamefont
  {Biscari}},\ }\bibfield  {title} {\enquote {\bibinfo {title} {The effect of
  defects on magnetic droplet nucleation},}\ }\href@noop {} {\bibfield
  {journal} {\bibinfo  {journal} {Physica A: Statistical Mechanics and its
  Applications}\ }\textbf {\bibinfo {volume} {611}},\ \bibinfo {pages} {128426}
  (\bibinfo {year} {2023})}\BibitemShut {NoStop}%
\bibitem [{\citenamefont {Yao}\ and\ \citenamefont
  {Jack}(2023)}]{yao_heterogeneous}%
  \BibitemOpen
  \bibfield  {author} {\bibinfo {author} {\bibfnamefont {L.}~\bibnamefont
  {Yao}}\ and\ \bibinfo {author} {\bibfnamefont {R.~L.}\ \bibnamefont {Jack}},\
  }\bibfield  {title} {\enquote {\bibinfo {title} {{Heterogeneous nucleation in
  the random field Ising model}},}\ }\href@noop {} {\bibfield  {journal}
  {\bibinfo  {journal} {The Journal of Chemical Physics}\ }\textbf {\bibinfo
  {volume} {159}},\ \bibinfo {pages} {244110} (\bibinfo {year}
  {2023})}\BibitemShut {NoStop}%
\bibitem [{\citenamefont {Poon}, \citenamefont {Seritan},\ and\ \citenamefont
  {Peters}(2015)}]{peters_additive_2015}%
  \BibitemOpen
  \bibfield  {author} {\bibinfo {author} {\bibfnamefont {G.~G.}\ \bibnamefont
  {Poon}}, \bibinfo {author} {\bibfnamefont {S.}~\bibnamefont {Seritan}},\ and\
  \bibinfo {author} {\bibfnamefont {B.}~\bibnamefont {Peters}},\ }\bibfield
  {title} {\enquote {\bibinfo {title} {A design equation for low dosage
  additives that accelerate nucleation},}\ }\href@noop {} {\bibfield  {journal}
  {\bibinfo  {journal} {Faraday Discuss.}\ }\textbf {\bibinfo {volume} {179}},\
  \bibinfo {pages} {329--341} (\bibinfo {year} {2015})}\BibitemShut {NoStop}%
\bibitem [{\citenamefont {Torrie}\ and\ \citenamefont
  {Valleau}(1977)}]{us_torrie_1977}%
  \BibitemOpen
  \bibfield  {author} {\bibinfo {author} {\bibfnamefont {G.}~\bibnamefont
  {Torrie}}\ and\ \bibinfo {author} {\bibfnamefont {J.}~\bibnamefont
  {Valleau}},\ }\bibfield  {title} {\enquote {\bibinfo {title} {Nonphysical
  sampling distributions in monte carlo free-energy estimation: Umbrella
  sampling},}\ }\href@noop {} {\bibfield  {journal} {\bibinfo  {journal} {J.
  Comput. Phys.}\ }\textbf {\bibinfo {volume} {23}},\ \bibinfo {pages}
  {187--199} (\bibinfo {year} {1977})}\BibitemShut {NoStop}%
\bibitem [{\citenamefont {Auer}\ and\ \citenamefont
  {Frenkel}(2004)}]{frenkel_2004_spherical_colloids}%
  \BibitemOpen
  \bibfield  {author} {\bibinfo {author} {\bibfnamefont {S.}~\bibnamefont
  {Auer}}\ and\ \bibinfo {author} {\bibfnamefont {D.}~\bibnamefont {Frenkel}},\
  }\bibfield  {title} {\enquote {\bibinfo {title} {Quantitative prediction of
  crystal-nucleation rates for spherical colloids: A computational approach},}\
  }\href@noop {} {\bibfield  {journal} {\bibinfo  {journal} {Annual Review of
  Physical Chemistry}\ }\textbf {\bibinfo {volume} {55}},\ \bibinfo {pages}
  {333--361} (\bibinfo {year} {2004})}\BibitemShut {NoStop}%
\bibitem [{\citenamefont {Allen}, \citenamefont {Valeriani},\ and\
  \citenamefont {ten Wolde}(2009)}]{2009_ffs_allen}%
  \BibitemOpen
  \bibfield  {author} {\bibinfo {author} {\bibfnamefont {R.~J.}\ \bibnamefont
  {Allen}}, \bibinfo {author} {\bibfnamefont {C.}~\bibnamefont {Valeriani}},\
  and\ \bibinfo {author} {\bibfnamefont {P.~R.}\ \bibnamefont {ten Wolde}},\
  }\bibfield  {title} {\enquote {\bibinfo {title} {Forward flux sampling for
  rare event simulations},}\ }\href@noop {} {\bibfield  {journal} {\bibinfo
  {journal} {J. Phys. Condens. Matter}\ }\textbf {\bibinfo {volume} {21}},\
  \bibinfo {pages} {463102} (\bibinfo {year} {2009})}\BibitemShut {NoStop}%
\bibitem [{\citenamefont {Escobedo}, \citenamefont {Borrero},\ and\
  \citenamefont {Araque}(2009)}]{2009_tps_escobedo}%
  \BibitemOpen
  \bibfield  {author} {\bibinfo {author} {\bibfnamefont {F.~A.}\ \bibnamefont
  {Escobedo}}, \bibinfo {author} {\bibfnamefont {E.~E.}\ \bibnamefont
  {Borrero}},\ and\ \bibinfo {author} {\bibfnamefont {J.~C.}\ \bibnamefont
  {Araque}},\ }\bibfield  {title} {\enquote {\bibinfo {title} {Transition path
  sampling and forward flux sampling. applications to biological systems},}\
  }\href@noop {} {\bibfield  {journal} {\bibinfo  {journal} {J. Phys. Condens.
  Matter}\ }\textbf {\bibinfo {volume} {21}},\ \bibinfo {pages} {333101}
  (\bibinfo {year} {2009})}\BibitemShut {NoStop}%
\bibitem [{\citenamefont {Allen}, \citenamefont {Warren},\ and\ \citenamefont
  {ten Wolde}(2005)}]{2005_ffs_allen}%
  \BibitemOpen
  \bibfield  {author} {\bibinfo {author} {\bibfnamefont {R.~J.}\ \bibnamefont
  {Allen}}, \bibinfo {author} {\bibfnamefont {P.~B.}\ \bibnamefont {Warren}},\
  and\ \bibinfo {author} {\bibfnamefont {P.~R.}\ \bibnamefont {ten Wolde}},\
  }\bibfield  {title} {\enquote {\bibinfo {title} {Sampling rare switching
  events in biochemical networks},}\ }\href@noop {} {\bibfield  {journal}
  {\bibinfo  {journal} {Phys. Rev. Lett.}\ }\textbf {\bibinfo {volume} {94}},\
  \bibinfo {pages} {018104} (\bibinfo {year} {2005})}\BibitemShut {NoStop}%
\bibitem [{\citenamefont {R\r{u}\v{z}i\v{c}ka}, \citenamefont {Quigley},\ and\
  \citenamefont {Allen}(2012)}]{polymer_folding_allen_2012}%
  \BibitemOpen
  \bibfield  {author} {\bibinfo {author} {\bibfnamefont {S.}~\bibnamefont
  {R\r{u}\v{z}i\v{c}ka}}, \bibinfo {author} {\bibfnamefont {D.}~\bibnamefont
  {Quigley}},\ and\ \bibinfo {author} {\bibfnamefont {M.~P.}\ \bibnamefont
  {Allen}},\ }\bibfield  {title} {\enquote {\bibinfo {title} {Folding kinetics
  of a polymer},}\ }\href@noop {} {\bibfield  {journal} {\bibinfo  {journal}
  {Phys. Chem. Chem. Phys.}\ }\textbf {\bibinfo {volume} {14}},\ \bibinfo
  {pages} {6044--6053} (\bibinfo {year} {2012})}\BibitemShut {NoStop}%
\bibitem [{\citenamefont {Blow}\ \emph {et~al.}(2023)\citenamefont {Blow},
  \citenamefont {Tribello}, \citenamefont {Sosso},\ and\ \citenamefont
  {Quigley}}]{blow_jcp_2023}%
  \BibitemOpen
  \bibfield  {author} {\bibinfo {author} {\bibfnamefont {K.~E.}\ \bibnamefont
  {Blow}}, \bibinfo {author} {\bibfnamefont {G.~A.}\ \bibnamefont {Tribello}},
  \bibinfo {author} {\bibfnamefont {G.~C.}\ \bibnamefont {Sosso}},\ and\
  \bibinfo {author} {\bibfnamefont {D.}~\bibnamefont {Quigley}},\ }\bibfield
  {title} {\enquote {\bibinfo {title} {{Interplay of multiple clusters and
  initial interface positioning for forward flux sampling simulations of
  crystal nucleation}},}\ }\href@noop {} {\bibfield  {journal} {\bibinfo
  {journal} {The Journal of Chemical Physics}\ }\textbf {\bibinfo {volume}
  {158}},\ \bibinfo {pages} {224102} (\bibinfo {year} {2023})}\BibitemShut
  {NoStop}%
\bibitem [{\citenamefont {Onsager}(1944)}]{onsager_1944}%
  \BibitemOpen
  \bibfield  {author} {\bibinfo {author} {\bibfnamefont {L.}~\bibnamefont
  {Onsager}},\ }\bibfield  {title} {\enquote {\bibinfo {title} {Crystal
  statistics. i. a two-dimensional model with an order-disorder transition},}\
  }\href@noop {} {\bibfield  {journal} {\bibinfo  {journal} {Phys. Rev.}\
  }\textbf {\bibinfo {volume} {65}},\ \bibinfo {pages} {117--149} (\bibinfo
  {year} {1944})}\BibitemShut {NoStop}%
\bibitem [{\citenamefont {Fletcher}(1958)}]{fletcher_1958}%
  \BibitemOpen
  \bibfield  {author} {\bibinfo {author} {\bibfnamefont {N.~H.}\ \bibnamefont
  {Fletcher}},\ }\bibfield  {title} {\enquote {\bibinfo {title} {{Size Effect
  in Heterogeneous Nucleation}},}\ }\href@noop {} {\bibfield  {journal}
  {\bibinfo  {journal} {The Journal of Chemical Physics}\ }\textbf {\bibinfo
  {volume} {29}},\ \bibinfo {pages} {572--576} (\bibinfo {year}
  {1958})}\BibitemShut {NoStop}%
\bibitem [{\citenamefont {Yu}(2003)}]{cross_nucleation1}%
  \BibitemOpen
  \bibfield  {author} {\bibinfo {author} {\bibfnamefont {L.}~\bibnamefont
  {Yu}},\ }\bibfield  {title} {\enquote {\bibinfo {title} {Nucleation of one
  polymorph by another},}\ }\href@noop {} {\bibfield  {journal} {\bibinfo
  {journal} {Journal of the American Chemical Society}\ }\textbf {\bibinfo
  {volume} {125}},\ \bibinfo {pages} {6380--6381} (\bibinfo {year}
  {2003})}\BibitemShut {NoStop}%
\bibitem [{\citenamefont {Looijmans}\ \emph {et~al.}(2018)\citenamefont
  {Looijmans}, \citenamefont {Cavallo}, \citenamefont {Yu},\ and\ \citenamefont
  {Peters}}]{cross_nucleation2}%
  \BibitemOpen
  \bibfield  {author} {\bibinfo {author} {\bibfnamefont {S.~F. S.~P.}\
  \bibnamefont {Looijmans}}, \bibinfo {author} {\bibfnamefont {D.}~\bibnamefont
  {Cavallo}}, \bibinfo {author} {\bibfnamefont {L.}~\bibnamefont {Yu}},\ and\
  \bibinfo {author} {\bibfnamefont {G.~W.~M.}\ \bibnamefont {Peters}},\
  }\bibfield  {title} {\enquote {\bibinfo {title} {Cross-nucleation between
  polymorphs: Quantitative modeling of kinetics and morphology},}\ }\href@noop
  {} {\bibfield  {journal} {\bibinfo  {journal} {Crystal Growth \& Design}\
  }\textbf {\bibinfo {volume} {18}},\ \bibinfo {pages} {3921--3926} (\bibinfo
  {year} {2018})}\BibitemShut {NoStop}%
\end{thebibliography}
%

\pagebreak
\beginsupplement

\widetext
\section*{Supplementary Information for "{\bf Mapping the influence of impurity interaction energy on nucleation in a lattice-gas model of solute precipitation}" }
\noindent {\bf{\textit{Authors:}} Dipanjan Mandal and David Quigley}

\noindent{\bf{\textit{Affiliation:}} Department of Physics, University of Warwick, Coventry CV4 7AL, United Kingdom}

\vspace{0.5cm}
\noindent In this supplementary information we provide additional data required to support the findings of the paper.

\subsection{\textit{Saturation of the free energy barrier and the nucleation rate:}}
\label{appdx:a}

\begin{figure}[ht!]
\centering
\includegraphics[width=0.45\columnwidth]{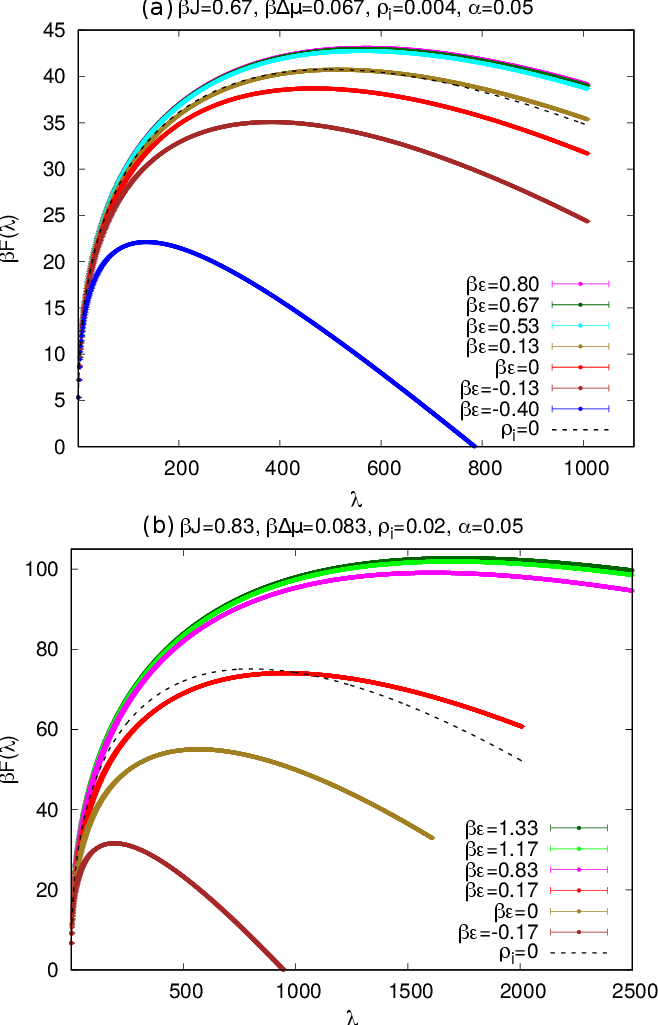}
\caption{Nucleation free energy with dynamic impurities, varying anti-symmetric interaction energy $\beta \epsilon_+=-\beta\epsilon_-=\beta\epsilon$, for (a) $\beta J=0.67$, $\beta \Delta\mu=0.067$, $\rho_i=0.004$ and (b) $\beta J=0.83$, $\beta \Delta\mu=0.083$, $\rho_i=0.02$ with fixed mobility parameter $\alpha=0.05$. Free energy barrier for the system without impurities ($\rho_i$=0) is plotted by black dotted line for comparison. The saturation in barrier height is seen for both low and high impurity density unlike the static impurities as shown in Fig.~\ref{fig:free_rhoi_p004}.}
\label{fig:free_sym_dyn}
\end{figure}

\begin{figure}[ht!]
\centering
\includegraphics[width=0.45\columnwidth]{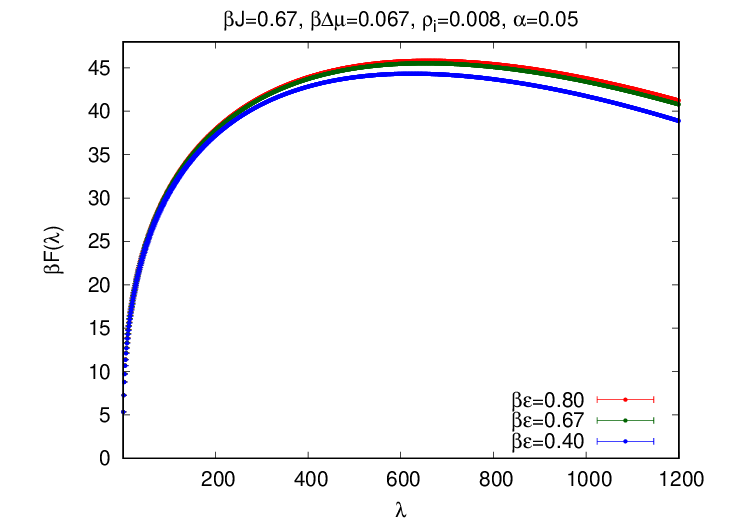}
\caption{Saturation of nucleation free energy barrier for anti-symmetric interaction energy $\beta\epsilon_+=-\beta\epsilon_-=\beta\epsilon$ with fixed $\beta J=0.67$, $\beta \Delta\mu=0.067$ in the presence of dynamic impurities of density $\rho_i=0.008$ with $\alpha=0.05$.}
\label{fig:free_sym_dyn2}
\end{figure}

As discussed in Section~4 of the paper, when impurities are dynamic and interaction energy is anti-symmetric ($\epsilon_+=-\epsilon_-=\epsilon$), the nucleation free energy barrier height and the nucleation rate do not depend on the interaction strength after a certain threshold value of the repulsive interaction energy $\epsilon_+$,  beyond this limit all impurities are removed from the cluster due to strong repulsive interaction with solute. We observe such saturation in free energy barrier with respect to the anti-symmetric interaction energy at both low $\rho_i=0.004$ and high $\rho_i=0.02$ impurity density (see Fig.~\ref{fig:free_sym_dyn}). Another example, for intermediate impurity density $\rho_i=0.008$, is shown in Fig.~\ref{fig:free_sym_dyn2} with parameter values $\beta J=0.67$, $\beta \Delta\mu=0.067$ and $\alpha=0.05$.

In the case of static impurities, we do not see such saturation in free energy barrier and nucleation rate above a certain impurity density threshold as discussed in Section~3 of the paper. For a random impurity configuration, the saturation criterion could be related with competition between the size of the largest void area without impurities and the critical cluster size. If the size of the largest void area is greater than the critical cluster size, we expect to see the saturation in the free energy barrier even for static impurities. Examining the detailed statistics of void site and distribution expected from a uniform distribution of impurities  could in principle lead to an estimate of that threshold. 
\begin{figure}[ht!]
\centering
\includegraphics[width=0.45\columnwidth]{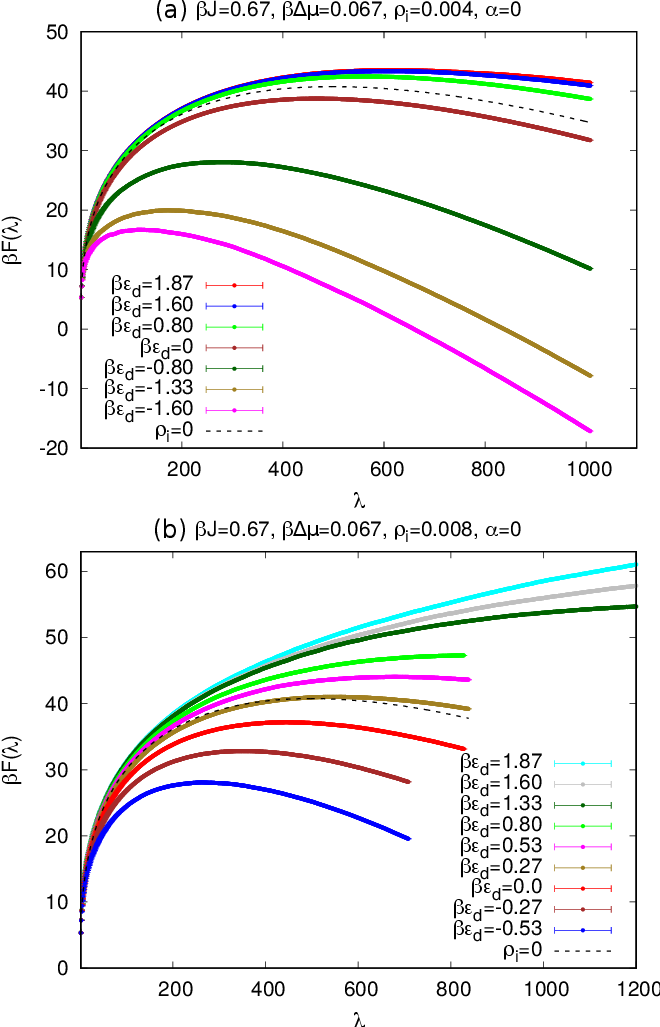}
\caption{Nucleation free energy with varying dimensionless interaction energy difference $\beta \epsilon_d$ with fixed $\beta J=0.67$, $\beta \Delta\mu=0.067$ for system size $L=100$ at static impurity density (a) $\rho_i=0.004$ and (b) $\rho_i=0.008$. We see no further increase in free energy barrier height with increasing $\beta \epsilon_d$ when $\rho_i=0.004$ or lower, i.e., when the impurities are sparsely distributed so that a critical cluster can fit in the void space without interacting with impurities. This behaviour in barrier height is not observed for $\rho_i=0.008$ when impurity density is higher. Free energy barrier for the system without impurities ($\rho_i$=0) is plotted by dotted line for comparison.}
\label{fig:free_rhoi_p004}
\end{figure}

\begin{figure}[ht!]
\centering
\includegraphics[width=0.45\columnwidth]{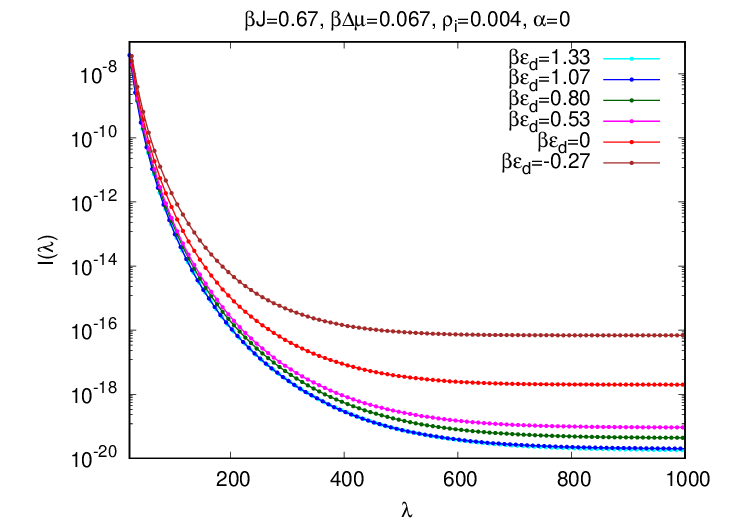}
\caption{Rate of obtaining a cluster of size $\lambda$ starting from a metastable solution phase for different interaction energy difference $\beta \epsilon_d$ at fixed $\beta J=0.67$, $\beta\Delta\mu=0.067$ and $\rho_i=0.004$ with static impurities. The constant value of $I(\lambda)$ for large $\lambda$ is the nucleation rate.}
\label{fig:rate_static_0.004}
\end{figure}
For the impurity density $\rho_i=0.004$, we observe saturation both in free energy barrier [see Fig.~\ref{fig:free_rhoi_p004}(a)] and nucleation rate [see Fig.~\ref{fig:rate_static_0.004}] with increasing $\beta\epsilon_d$. However, at higher impurity density $\rho_i=0.008$, when the average void area excluding impurities decreases, we observe a monotonic increase in barrier height without saturation as $\beta\epsilon_d$ is increased [see Fig.~\ref{fig:free_rhoi_p004}(b)]. 
We also note the differences in shape of saturated free energy between static and dynamic cases at $\rho_i=0.004$. Unlike dynamic impurities, the free energy curve becomes flatter in the case of static impurities and starts to deviate from the standard form of the free energy function assumed in classical nucleation theory as given in Eq.4 (see $\beta \epsilon_d=1.6$ curve in Fig.\ref{fig:bd_rhoi0.004_st} and high positive $\beta \epsilon_d$ curves in Fig.~\ref{fig:free_rhoi_p004}). The confinement/constraint imposed by the immobile impurities could be responsible for this behaviour as it forces nuclei into shape with surface area to perimeter ratios that differ from the ideal case.

\subsection{\textit{Free energy barrier with symmetric interaction energy ($\epsilon_+=\epsilon_-=\epsilon$):}}
\label{appdx:a1}
\begin{figure}[ht!]
\centering
\includegraphics[width=0.45\columnwidth]{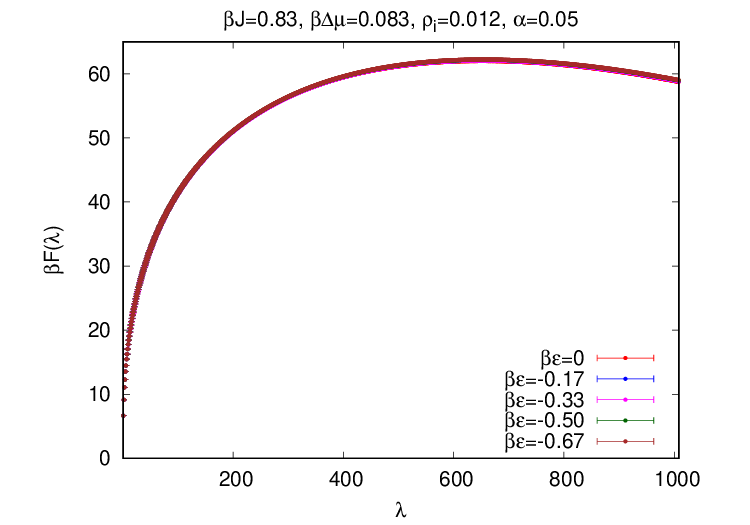}
\caption{Nucleation free energy with dynamic impurities, varying symmetric interaction strengths $\beta \epsilon_+=\beta \epsilon_-=\beta \epsilon$ for $\beta J=0.83$, $\beta \Delta\mu=0.083$,  $\rho_i=0.012$ and $\alpha=0.05$. The plotted range of interaction energy lie in {\it{surfactant}} regime of the behaviour map. We do not see any variation in barrier height.}
\label{fig:free_sym_rhoi0.012}
\end{figure}
In the {\it{surfactant}} regime of the behaviour map we do not see much variation in the nucleation rate as shown in Fig.~8(a) of the paper, for dynamic impurities with $\beta J=0.83$, $\beta \Delta\mu=0.083$, $\rho_i=0.012$ and $\alpha=0.05$. Similar behaviour is reflected in free energy plots for different values of symmetric interaction energies that belong to the {\it{surfactant}} regime as shown in Fig.~\ref{fig:free_sym_rhoi0.012}.

\subsection{\textit{Decrement in free energy barrier height due to mobile impurities:}}
\label{appdx:b}
\begin{figure}[ht!]
\centering
\includegraphics[width=0.45\columnwidth]{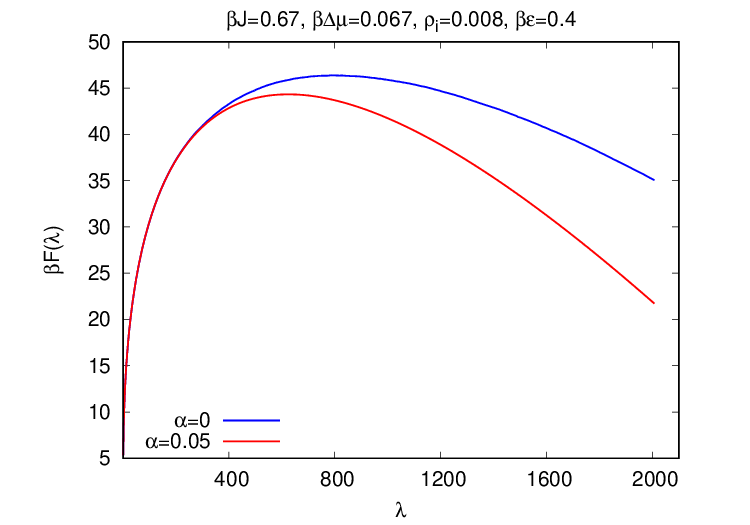}
\caption{Comparison of the free energy barrier for static ($\alpha=0$) and dynamic ($\alpha=0.05$) impurities with same anti-symmetric interaction energy $\beta \epsilon_+=-\beta\epsilon_-=0.4$ for $\beta J=0.67$ with $\rho_i=0.008$.}
\label{fig:compare_static_dyn}
\end{figure}

As observed in Section~4, the free energy barrier height to nucleation decreases when impurities are dynamic compared to the static case for same set of interaction energies as shown in Fig.~\ref{fig:compare_static_dyn} (see Fig. 2(a) of the paper). In this case the interaction energies are anti-symmetric ($\beta\epsilon_+=-\beta\epsilon_-=0.4$) and other parameter values are $\beta J=0.67$, $\beta \Delta\mu=0.067$ and $\rho_i=0.008$.  Dynamic impurities enhance the nucleation rate by decreasing the barrier height. In this example the microscopic interaction between impurity-solute  and impurity-solvent are respectively weakly-repulsive and weakly-attractive and lies at the boundary of the {\it{surfactant}} regime of the behaviour map. We also observed similar decrement in barrier height in our earlier work [D. Mandal and D. Quigley, Soft Matter, 2021, 17, 8642–8650] for impurities with neutral interactions. However, in that case, because of the neutral interaction impurities prefer to stay at the boundaries of the cluster, it reduces the interfacial free energy.

\subsection{\textit{Fitting free energy barrier:}}
\label{appdx:c}
\begin{figure}[ht!]
\centering
\includegraphics[width=0.45\columnwidth]{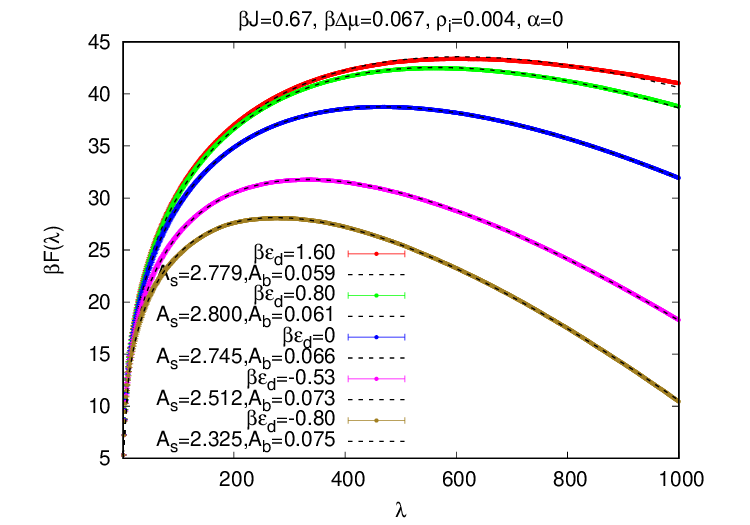}
\caption{Fitting of the free energy expression given in Eq.~4 with the free energy obtained from umbrella sampling method varying $\beta\epsilon_d$ for static impurities with $\beta J=0.67$, $\beta\Delta\mu =0.067$ and $\rho_i=0.004$.}
\label{fig:bd_rhoi0.004_st}
\end{figure}

Fitting of free energy to the expression given in Eq.~4 in the case of static impurities with density $\rho_i=0.004$, $\beta J=0.67$ and $\beta \Delta\mu=0.067$ for different $\beta \epsilon_d$ is shown in Fig.~\ref{fig:bd_rhoi0.004_st}, where we allow  the surface $A_s$ and bulk $A_b$ terms to vary from the $\rho_i=0$ case. We see that $A_s$ increases and $A_b$ decreases monotonically with increasing $\beta \epsilon_d$ from negative to positive values. The fitting becomes more accurate with decreasing $\beta \epsilon_d$. 

\begin{figure}[ht!]
\centering
\includegraphics[width=0.45\columnwidth]{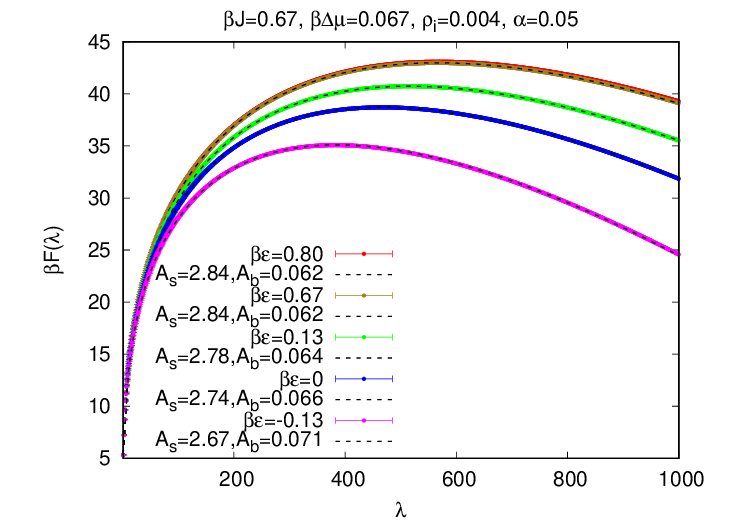}
\caption{Fitting of the free energy expression given in Eq.~4 with the free energy obtained from umbrella sampling method for dynamic impurities and varying anti-symmetric interaction energy $\beta\epsilon_+=-\beta\epsilon_-=\beta\epsilon$ with $\beta J=0.67$, $\beta\Delta\mu =0.067$, $\rho_i=0.004$ and $\alpha=0.05$.}
\label{fig:bd_rhoi0.004_dyn}
\end{figure}

Similar fitting of calculated free energy with Eq.~4 for dynamic impurities with $\beta J=0.67$, $\beta \Delta\mu=0.067$, $\rho_i=0.004$ and $\alpha=0.05$  is shown in Fig.~\ref{fig:bd_rhoi0.004_dyn} for different $\beta \epsilon$. We see monotonic decrease and monotonic increase of the bulk and surface terms respectively with increasing $\beta \epsilon$ until they converge to non-zero finite values after entering into the inert-spectator regime of the behaviour map. We note that, for pure Ising model at low temperatures $\Delta g\approx\Delta\mu$, and from Fig.~\ref{fig:bd_rhoi0.004_dyn} we see that this relation holds for neutral impurities, as $A_b\approx\beta \Delta\mu$ when $\beta \epsilon=0$, but not for non-zero interaction energies.

\subsection{\textit{Nucleation rate and diffusion coefficient $D_c$:}}
\label{appdx:d}
\begin{figure}[ht!]
\centering
\includegraphics[width=0.45\columnwidth]{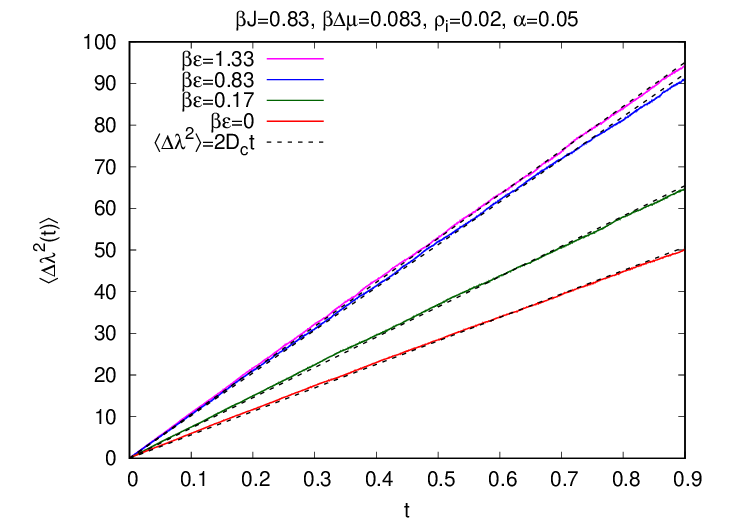}
\caption{Linear fitting of diffusivity $\langle \Delta \lambda^2\rangle=2D_ct$ obtained from simulation for dynamic impurities with $\beta J=0.83$, $\beta \Delta\mu=0.083$, $\rho_i=0.02$ and $\alpha=0.05$. Initial size of the cluster is set to the critical cluster size calculated from the position of the maxima in the respective $F(\lambda)$ vs. $\lambda$ plots.}
\label{fig:diff_T1.2imp0.02}
\end{figure}
The Becker-Doring-Zeldovitch nucleation rate $I_{BDZ}$ is calculated using Eq.~5 of the paper for high $\rho_i=0.02$ and intermediate $\rho_i=0.012$ impurity densities. For that, the diffusion coefficient $D_c$ is calculated after performing independent simulations starting from the critical cluster size at time $t=0$ and calculating the slope of the mean squared deviation of cluster size with time $t$. Estimated values of $D_c$ corresponding to plots displayed in Fig.~\ref{fig:diff_T1.2imp0.02} are written in column 4 of Table~\ref{tab:two}. The values of different parameters required for calculating $I_{BDZ}$ is given in Table~\ref{tab:two} and Table~\ref{tab:three} for $\rho_i=0.02$ (anti-symmetric interaction energy) and $\rho_i=0.012$ (symmetric interaction energy) respectively. In the final column the nucleation rate obtained from independent forward flux sampling simulations $I_{FFS}$ is written. The results for $I_{BDZ}$ and $I_{FFS}$ match quite well for the range of interaction energies considered in both tables.

\begin{table}
\begin{center}
\begin{tabular}{|c c c c c c|}
 \hline
 $\beta \epsilon$ & $\lambda_c$ & $\beta F(\lambda_c)$ & $D_c$ & $I_{BDZ}$ & $I_{FFS}$ \\ [0.5ex] 
 \hline\hline
 $1.33$ & $1689$ & $103.58$ & $52.8$ & $9\times10^{-47}$ & $1.7\times10^{-46}$ \\ [0.5ex] 
 \hline
 $0.83$ & $1584$ & $99.26$ & $51.3$ & $6.8\times10^{-45}$ & $5.3\times10^{-45}$  \\ [0.5ex] 
 \hline
 $0.17$ & $942$ & $73.69$ & $36.4$ & $8.8\times10^{-34}$ & $5.6\times10^{-34}$  \\ [0.5ex] 
 \hline
 $0$ & $571$ & $55.21$ & $28.2$ & $1\times10^{-25}$ & $1.1\times10^{-25}$  \\ [0.5ex] 
 \hline
\end{tabular}
 \caption{Comparison of nucleation rates obtained from Becker-Doring-Zeldovich analysis ($I_{BD}$) and forward flux sampling method ($I_{FFS}$) for $\beta J=0.83$, $\rho_i=0.02$ and $\alpha=0.05$ with anti-symmetric interaction energy $\beta\epsilon_+=-\beta\epsilon_-=\epsilon$. The maximum error in determining $I_{BDZ}$ and $I_{FFS}$ are $80\%$ and $10\%$ respectively. The parameter values $\beta\epsilon=0$ and $\beta\epsilon=1.33$ lie in the surfactant and bulk-stabilizer regimes respectively.}
  \label{tab:two}
  \end{center}
\end{table}

\begin{table}
\begin{center}
\begin{tabular}{|c c c c c c|}
 \hline
 $\beta \epsilon$ & $\lambda_c$ & $\beta F(\lambda_c)$ & $D_c$ & $I_{BDZ}$ & $I_{FFS}$ \\ [0.5ex] 
 \hline\hline
 $0.67$ & $642$ & $60.88$ & $30.4$ & $3.5\times10^{-28}$ & $7.8\times10^{-28}$ \\ [0.5ex] 
 \hline
 $1.0$ & $670$ & $56.93$ & $30.6$ & $1.8\times10^{-26}$ & $2.5\times10^{-26}$  \\ [0.5ex] 
 \hline
 $1.33$ & $766$ & $61.58$ & $33.5$ & $1.6\times10^{-28}$ & $6.7\times10^{-28}$  \\ [0.5ex] 
 \hline
\end{tabular}
 \caption{Comparison of nucleation rates obtained from Becker-Doring-Zeldovich analysis ($I_{BDZ}$) and forward flux sampling method ($I_{FFS}$) for $\beta J=0.83$, $\beta \Delta\mu=0.083$, $\rho_i=0.012$ and $\alpha=0.05$ with symmetric interaction energy $\beta\epsilon_+=\beta\epsilon_-=\epsilon$. This range of $\beta\epsilon$ belongs to the regime in which impurities act like heterogeneous nucleating sites.}
  \label{tab:three}
  \end{center}
\end{table}
\end{document}